\documentclass[preprint]{aastex}
\usepackage{hyperref}

\begin{document}

\title{CO~($J = 1-0$) Observations of a Filamentary Molecular Cloud in the Galactic Region Centered at $l = 150\arcdeg, b = 3.5\arcdeg$}

\author{Fang Xiong\altaffilmark{1,2}, Xuepeng Chen\altaffilmark{1}, Ji Yang\altaffilmark{1}, Min Fang\altaffilmark{1}, Shaobo Zhang\altaffilmark{1}, Miaomiao Zhang\altaffilmark{1},\\ Xinyu Du\altaffilmark{1,2}, and Wenshan Long\altaffilmark{1,3}}

\affil{$^1$Purple Mountain Observatory, and Key Laboratory of Radio Astronomy, Chinese Academy of Sciences, 2 West Beijing Road, Nanjing 210008, China; xpchen@pmo.ac.cn}
\affil{$^2$University of Chinese Academy of Sciences, 19A Yuquan Road, Shijingshan District, Beijing 100049, China; fangxiong@pmo.ac.cn}
\affil{$^3$Shanghai Normal University, 100 Guilin Road, Xuhui District, Shanghai 200234, China}

\begin{abstract}
We present large-field (4.25~$\times$~3.75 deg$^2$) mapping observations toward the Galactic region centered at $l = 150\arcdeg, b = 3.5\arcdeg$ in the $J = 1-0$ emission line of CO isotopologues ($^{12}$CO, $^{13}$CO, and C$^{18}$O), using the 13.7 m millimeter-wavelength telescope of the Purple Mountain Observatory. Based on the $^{13}$CO observations, we reveal a filamentary cloud in the Local Arm at a velocity range of $-$0.5 to 6.5~km~s$^{-1}$. This molecular cloud contains 1 main filament and 11 sub-filaments, showing the so-called ``ridge-nest'' structure. The main filament and three sub-filaments are also detected in the C$^{18}$O line. The velocity structures of most identified filaments display continuous distribution with slight velocity gradients. The measured median excitation temperature, line width, length, width, and linear mass of the filaments are $\sim$9.28~K, 0.85~km~s$^{-1}$, 7.30~pc, 0.79~pc, and 17.92~$M_\sun$~pc$^{-1}$, respectively, assuming a distance of 400~pc. We find that the four filaments detected in the C$^{18}$O line are thermally supercritical, and two of them are in the virialized state, and thus tend to be gravitationally bound. We identify in total 146 $^{13}$CO clumps in the cloud, about 77$\%$ of the clumps are distributed along the filaments. About 56$\%$ of the virialized clumps are found to be associated with the supercritical filaments. Three young stellar object (YSO) candidates are also identified in the supercritical filaments, based on the complementary infrared (IR) data. These results indicate that the supercritical filaments, especially the virialized filaments, may contain star-forming activities.
\end{abstract}


\section{INTRODUCTION}
Filamentary structures are frequently seen in nearby star-forming regions (SFRs), such as the Orion cloud \citep{bal87, chi97}, the Taurus cloud \citep{abe94, miz95}, the Ophiuchus cloud \citep{lor89}, and the Perseus cloud \citep{hat05}, and have also been revealed in the distant infrared dark clouds \citep[IRDCs; e.g.,][]{per09} and cold interstellar medium \citep[ISM; e.g.,][]{and10, men10, mol10}. The ubiquity of filamentary structures suggests that they exist such a long time during the lifetime of clouds that they can provide important information about the origin of SFRs \citep{mye09}. Numerous studies indicate that filamentary structures represent a key step in the process of star formation, which connects the compression of molecular gas with the fragmentation into prestellar cores \citep[see a recent review by][and references therein]{and14}. However, the nature of filamentary structures is still unclear \citep{mye09, and14}, and their spatial organization is also under controversy.

Filaments are typically elongated structures, with the lengths ranging from a few parsecs in the nearby molecular clouds to several tens of parsecs in some IRDCs \citep{beu11, jac10}. \citet{arz11} found that filaments in the Gould Belt have a narrow distribution of width with a median value of 0.10~$\pm$~0.03~pc. With the length and width of filaments, \citet{and14} defined a filament as an elongated structure with an aspect ratio larger than $\sim$5 -- 10 that is significantly overdense compared to its surrounding ISM. They also pointed out that filaments are generally linear over their length and appear to be co-linear in the direction of the longer extents of their host clouds.

For the spatial organization of filaments, \citet{tac02} suggested the so-called ``head-tail'' structure, in which ``head'' is the central region where cluster formation takes place and ``tail'' is the filamentary extension of the central region. \citet{mye09} noticed that some ``heads'' have more than one associated ``tail'' in the deeper and higher angular resolution observations, and presented the ``hub-filament'' structure. In this model, ``hub'' is the central body of low aspect ratio and high column density, while ``filaments'' are the associated features of higher aspect ratio and lower column density. \citet{hil11} found that the filaments in the Vela C molecular cloud complex appear to be more uni-directional within ``ridges'' in inner, higher column density area and more variedly directional within ``nests'' in outer, lower column density area. The magnetic field appears to be perpendicular to the ``ridges'' and chaotic in the ``nests'' \citep{kus16}. In the Taurus molecular cloud, on the other hand, \citet{pal13} found that the dense, star-forming filament B211 is surrounded by a large number of low-density sub-filaments (so-called ``striations'') oriented roughly perpendicular to the main filament, along with the magnetic field running parallel to the ``striations'' and perpendicular to the main filament. Moreover, as pointed out by \citet{pal13} and \citet{kus16}, the magnetic field plays a significant role in shaping the morphology of the filaments in both ``ridge-nest'' and ``filament-striation'' structures.

It is of importance to search for more filamentary structures and study their properties in detail, as well as star-forming activities therein, in order to achieve a better understanding of the nature of filaments. The Columbia-CfA $^{12}$CO~($J = 1-0$) line survey toward the Galactic Plane\footnote{\url{http://www.cfa.harvard.edu/mmw/MilkyWayinMolClouds.html}} \citep[see][]{dam87, dam01} had found a large elongated molecular cloud in the Galactic region centered at $l = 150\arcdeg, b = 3.5\arcdeg$ (referred to as the ``G150 region'' hereafter). This elongated cloud is located to the west of a giant molecular cloud (GMC), and it covers an area within $148\arcdeg \leq l \leq 153\arcdeg$ and $1\arcdeg \leq b \leq 6\arcdeg$ (see Figure~\ref{fig:fig1}). A few supernova remnants (SNRs) identified by radio observations are also located in this area, like G149.5+3.2 and G150.8+3.8 \citep{ger14}, and G150.3+4.5 \citep{gao14}. Three velocity components can be roughly resolved in the longitude-velocity map of this elongated cloud (see Figure~\ref{fig:fig2}), with two GMCs located in the east and west. Nevertheless, there is no detailed study toward this elongated cloud thus far and the properties of molecular gas in the cloud still remain unknown.

In this paper, we present new CO~($J = 1-0$) observations toward the G150 region in the $^{12}$CO, $^{13}$CO, and C$^{18}$O lines, using the 13.7 m telescope of the Purple Mountain Observatory (PMO), which is part of the Milky Way Imaging Scroll Painting (MWISP) project for investigating the molecular gas along the northern Galactic Plane. Based on the high-resolution CO multi-line observations (the angular resolution in the MWISP survey is $\sim$50$\arcsec$, while the angular resolution in the Columbia-CfA survey toward the second Galactic quadrant is 0.25$\arcdeg$), we resolve the elongated G150 molecular gas into a large filamentary cloud, and study in detail the properties of this filamentary cloud in this work. In Section~\ref{sec:observations}, we describe the CO line observations and data reduction. Observational results are presented in Section~\ref{sec:results} and discussed in Section~\ref{sec:discussion}. The main conclusions of this work are summarized in Section~\ref{sec:summary}.

\section{OBSERVATIONS AND DATA REDUCTION}\label{sec:observations}

\subsection{CO Data}
Our observation of the G150 region was conducted by the 13.7 m millimeter-wavelength telescope of PMO located in Delingha, China, from 2013 September to 2014 November. The whole observed region covered an area within $147.75\arcdeg \leq l \leq 152\arcdeg$ and $1.5\arcdeg \leq b \leq 5.25\arcdeg$, and was observed in $^{12}$CO~($J = 1-0$), $^{13}$CO~($J = 1-0$), and C$^{18}$O~($J = 1-0$) simultaneously with the 9-beam Superconducting Spectroscopic Array Receiver \citep{sha12}. Using the on-the-fly (OTF) observation mode, the telescope scanned the sky along both longitude and latitude directions at a constant rate of 50$\arcsec$~s$^{-1}$, and the receiver recorded at an interval of 0.3~s.

The half power beam width of the telescope is about 52$\arcsec$ at 115.2~GHz, and 50$\arcsec$ at 110.2~GHz. The antenna temperature ($T_{\rm A}^*$) is calibrated to the main beam temperature ($T_{\rm MB}$) by using $T_{\rm MB} = T_{\rm A}^*/\eta_{\rm MB}$, with the main beam efficiency ($\eta_{\rm MB}$) of 44$\%$ for $^{12}$CO and 48$\%$ for $^{13}$CO and C$^{18}$O. The typical system temperature is 270~K for $^{12}$CO, and 180~K for $^{13}$CO and C$^{18}$O during the observation.

After deriving the original OTF data, we checked and removed the bad channels and standing waves in the spectra. Then we regridded the checked data to $30\arcsec \times 30\arcsec$ pixels and converted them into the standard FITS files by the GILDAS software package \citep{gui00}. Using the Interactive Data Language software package with astronomy library, we mosaicked these FITS files together. In the resulting data cube, the rms noise level was 0.48~K for $^{12}$CO at a velocity resolution of 0.16~km~s$^{-1}$, and 0.28~K for $^{13}$CO and C$^{18}$O at a velocity resolution of 0.17~km~s$^{-1}$.

\subsection{Infrared Data}
The complementary IR data used in this work were obtained from the Two Micron All Sky Survey \citep[2MASS,][]{skr06} and the Wide-field Infrared Survey Explorer \citep[WISE,][]{wri10}. 2MASS collected the photometry over the entire celestial sphere in the near-IR $J$ (1.25~$\mu$m), $H$ (1.65~$\mu$m), and $K_{\rm s}$ (2.16~$\mu$m) bands with 10~$\sigma$ point-source detection levels of $\sim$15.8, 15.1, and 14.3~mag, respectively. WISE mapped the whole sky in four IR bands $W1$ (3.4~$\mu$m), $W2$ (4.6~$\mu$m), $W3$ (12~$\mu$m), and $W4$ (22~$\mu$m) with 5~$\sigma$ point-source sensitivities of 0.08, 0.11, 1, and 6~mJy, respectively. The IR data were retrieved from the NASA/IPAC Infrared Science Archive (IRSA)\footnote{\url{http://irsa.ipac.caltech.edu/frontpage/}}.

\section{RESULTS}\label{sec:results}

\subsection{Overview of the G150 Region}\label{sec:overview}
\begin{figure}
\epsscale{1.0}
\plotone{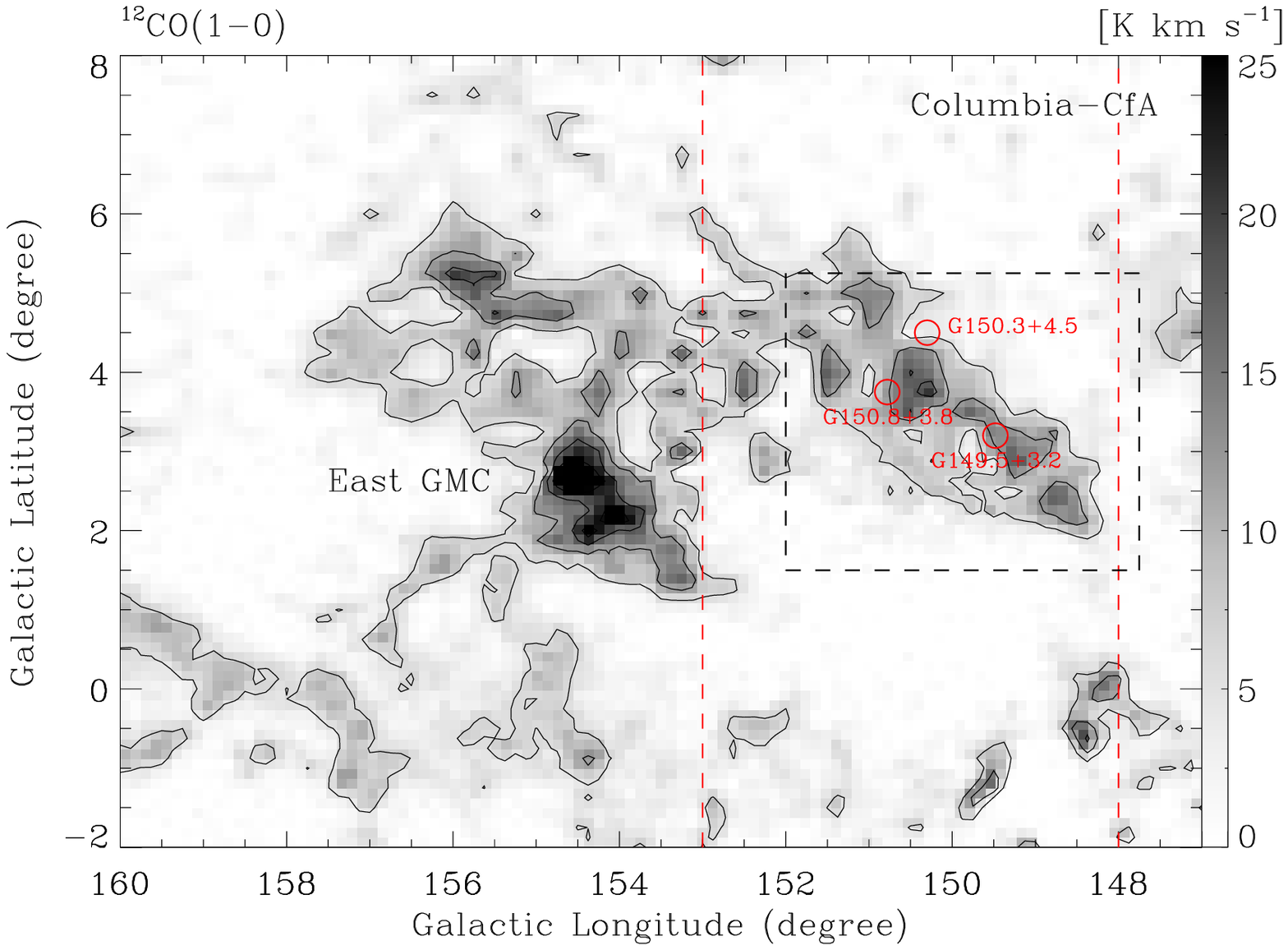}
\caption{Integrated intensity map of the Columbia-CfA $^{12}$CO~($J = 1-0$) Survey toward the Galactic Plane within $147\arcdeg \leq l \leq 160\arcdeg$ and $-2\arcdeg \leq b \leq -8\arcdeg$. The integrated velocity range is from $-$15 to 8.5~km~s$^{-1}$. The contours are 20$\%$, 40$\%$, 60$\%$, and 80$\%$ of the maximum value ($\sim$30~K) of the $^{12}$CO emission. The vertical dashed red lines mark the longitude range of this elongated molecular cloud, which corresponds to the vertical dashed red lines in Figure~\ref{fig:fig2}. The red circles indicate the positions of identified SNRs (see the text). The dashed rectangle shows the area of CO observations in this work, which corresponds to the dashed rectangle in Figure~\ref{fig:fig2}.}
\label{fig:fig1}
\end{figure}

\begin{figure}
\epsscale{1.0}
\plotone{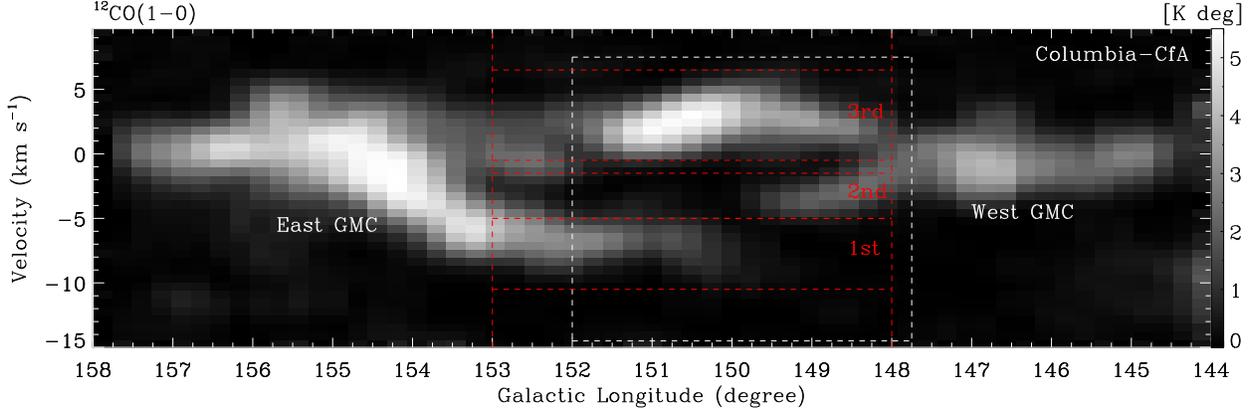}
\caption{Longitude-velocity map of the Columbia-CfA $^{12}$CO~($J = 1-0$) Survey toward the Galactic Plane within $144\arcdeg \leq l \leq 158\arcdeg$ and $-$15~km~s$^{-1}$ $\leq v \leq$ 10~km~s$^{-1}$. The integrated latitude range is from 1$\arcdeg$ to 6$\arcdeg$. The vertical dashed red lines mark the longitude range of the elongated molecular cloud shown in Figure~\ref{fig:fig1}. The horizontal dashed red lines indicate the ranges of three velocity components. The dashed rectangle shows the area of CO observations in this work.}
\label{fig:fig2}
\end{figure}

\begin{figure}
\epsscale{1.0}
\plotone{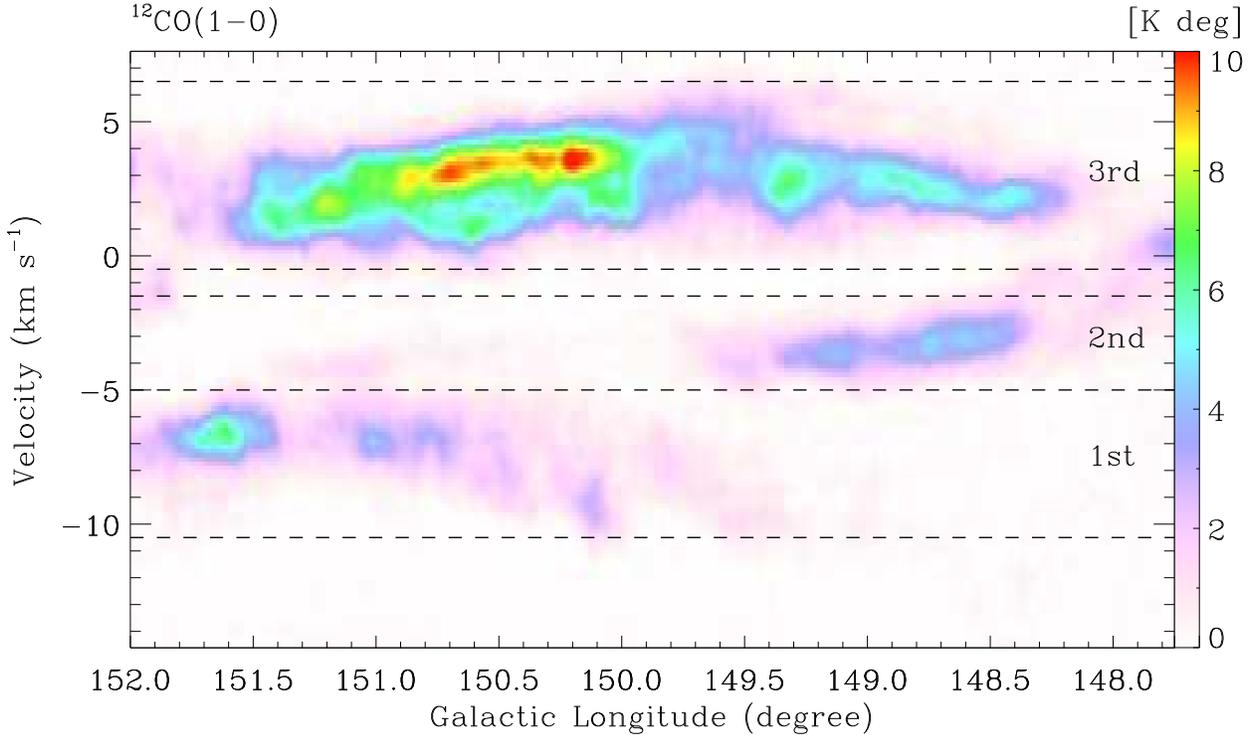}
\caption{Longitude-velocity map of the G150 region in the $^{12}$CO emission. The integrated latitude range is from 1.5$\arcdeg$ to 5.25$\arcdeg$. The dashed lines indicate the ranges of the velocity components, which are $-$10.5 to $-$5~km~s$^{-1}$, $-$5 to $-$1.5~km~s$^{-1}$, and $-$0.5 to 6.5~km~s$^{-1}$.}
\label{fig:fig3}
\end{figure}

\begin{figure}
\epsscale{1.0}
\plotone{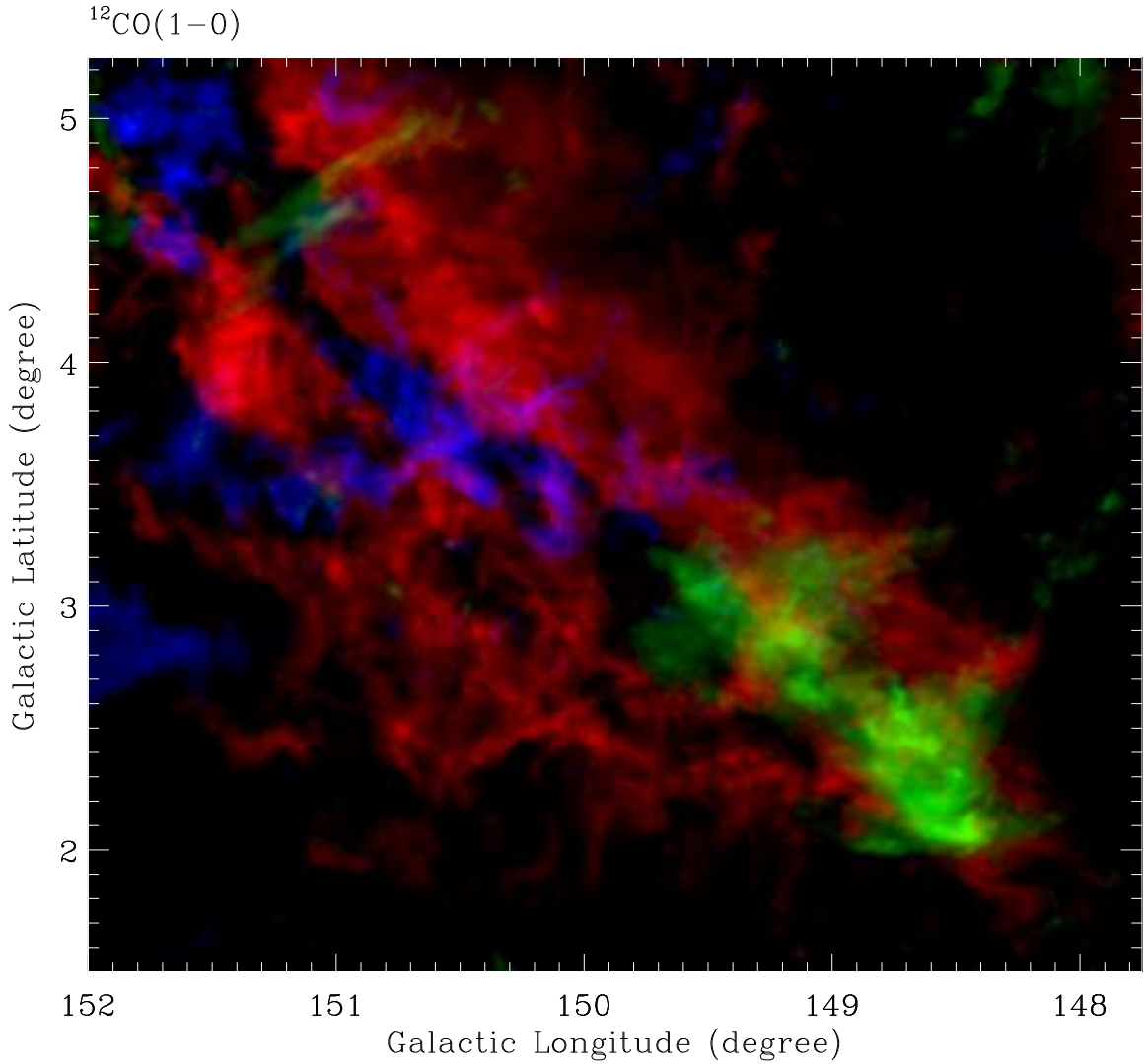}
\caption{Three-color image of velocity components in the $^{12}$CO emission. The blue area is the first velocity component, the green area is the second velocity component, and the red area is the third velocity component.}
\label{fig:fig4}
\end{figure}

As shown in Figure~\ref{fig:fig1} and Figure~\ref{fig:fig2}, the area of our CO observations of the G150 region has covered the main part of the elongated molecular cloud found by the Columbia-CfA survey. Figure~\ref{fig:fig3} presents our result of the longitude-velocity map of the G150 region. Three velocity components are clearly resolved with the ranges of $-$10.5 to $-$5~km~s$^{-1}$ (first velocity component), $-$5 to $-$1.5~km~s$^{-1}$ (second velocity component), and $-$0.5 to 6.5~km~s$^{-1}$ (third velocity component), respectively. The first component is distributed from 149.5$\arcdeg$ to 152$\arcdeg$ in the direction of longitude, while the second component is distributed from 148$\arcdeg$ to 149.5$\arcdeg$. Compared to the former two components, the third component, which is distributed from 148$\arcdeg$ to 152$\arcdeg$, is more extended in the direction of longitude and has higher intensity of the $^{12}$CO emission. The spatial distribution of these three velocity components in the $^{12}$CO emission is shown in Figure~\ref{fig:fig4}. The first component (blue area) is mainly distributed in the northeast of the G150 region, while the second component (green area) is in the southwest. The third component (red area) is distributed from northeast to southwest, covering a much larger area and presenting the shape of a large filament.

According to the results of \citet{dam01}, the velocity range of the Local Arm in the direction of G150 is roughly from $-$15 to 10~km~s$^{-1}$, which means these velocity components should be located in the Local Arm. We further calculate the distances of these components to confirm this idea. For the former two components, we derive the heliocentric distances of 340~pc (first component) and 60~pc (second component) using the spatial-kinematic method based on the galactic parameters of model A5 in \citet{rei14}. For the last component, we adopt a method based on near-infrared photometry (see Section~\ref{sec:distance}) to derive a heliocentric distance of 400~pc. Then, according to the characteristics of spiral arms in the Milky Way \citep[Table 2 of][]{rei14} and the galactocentric radius of the Sun \citep[8.34~$\pm$~0.16~kpc in][]{rei14}, we find that these three components are indeed located in the Local Arm.

Shown in Figure~\ref{fig:fig2}, the first and second components seem to be connected to the East GMC and West GMC, respectively, while the third component looks like a ``bridge'' between the East GMC and West GMC. Along with the spatial distribution shown in Figure~\ref{fig:fig4}, we suggest that these components may be different layers belonging to different GMCs in the Local Arm. Regarding the formation and dynamical interaction of these components, we speculate that these components and GMCs are used to be an entirety, thus a larger GMC, and separate from each other under the internal motions of the larger GMC (e.g., rotation, expansion). Another possible speculation is that these two GMCs are used to be different inhomogeneous clouds, and these three components, including the filamentary structures, are generated in the shocked layer during the collision of these two inhomogeneous GMCs, as suggested by isothermal MHD simulations \citep{ino13}. However, the resolution of the Columbia-CfA survey is not good enough to reveal the details of these clouds in Figure~\ref{fig:fig2}, and our observations have not covered the entire region in Figure~\ref{fig:fig1} and Figure~\ref{fig:fig2} so far. A future large-scale CO survey with high resolution will help us to develop a better understanding of the physical relation of these components and GMCs. In this work, we tend to regard these three velocity components as different gas layers in the Local Arm.

\begin{figure}
\epsscale{1.0}
\plotone{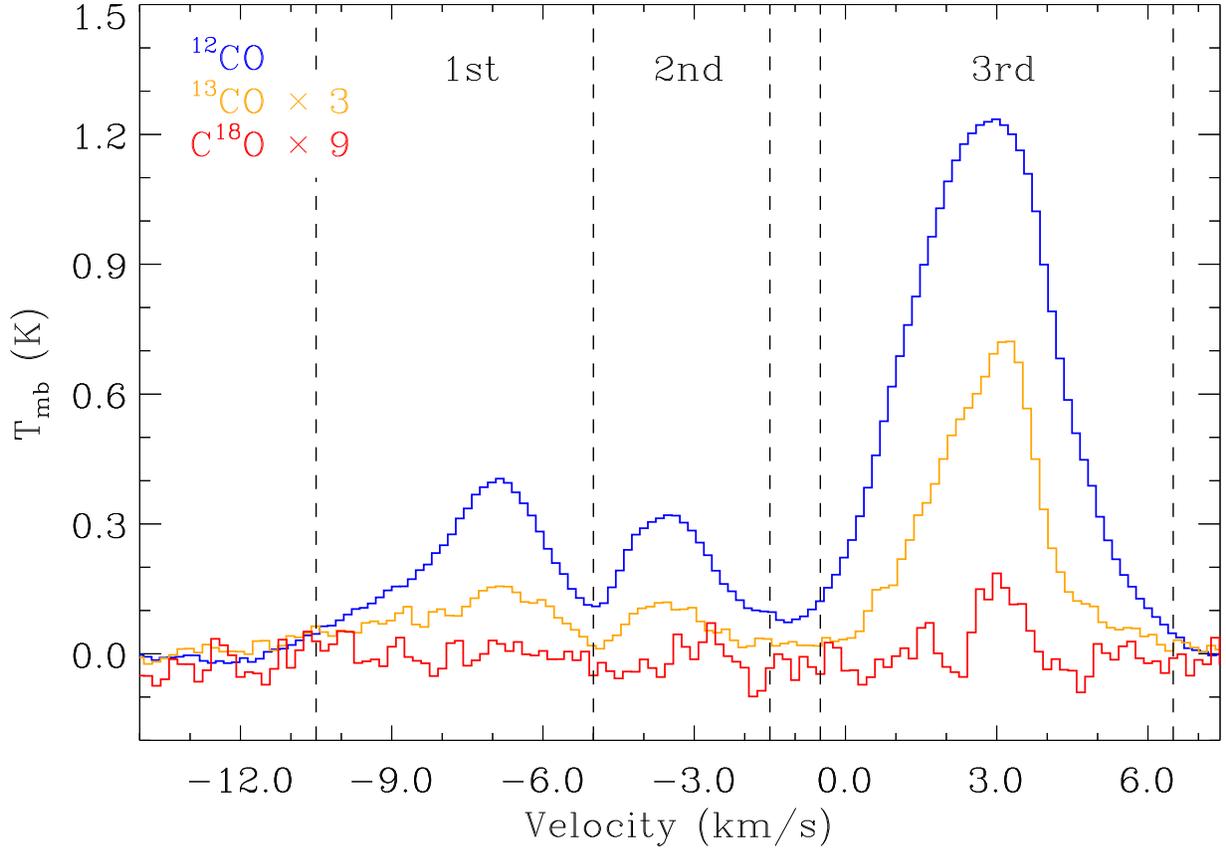}
\caption{CO spectra averaged in the whole G150 region. The blue spectrum shows the $^{12}$CO emission, the orange spectrum shows the $^{13}$CO emission multiplied by a factor of three, and the red spectrum shows the C$^{18}$O emission multiplied by a factor of nine. The dashed lines indicate the ranges of three velocity components.}
\label{fig:fig5}
\end{figure}

Based on the CO observations, we investigate the basic physical properties of these three velocity components. In addition to the $^{12}$CO emission, we also calculate the $^{13}$CO and C$^{18}$O emission of these components, which are presented in Figure~\ref{fig:fig5}. The CO spectra show that the $^{12}$CO and $^{13}$CO emission of the third component are much stronger than those of the former two components, and the C$^{18}$O emission is more significant in the third component.

\begin{figure}
\epsscale{1.0}
\plotone{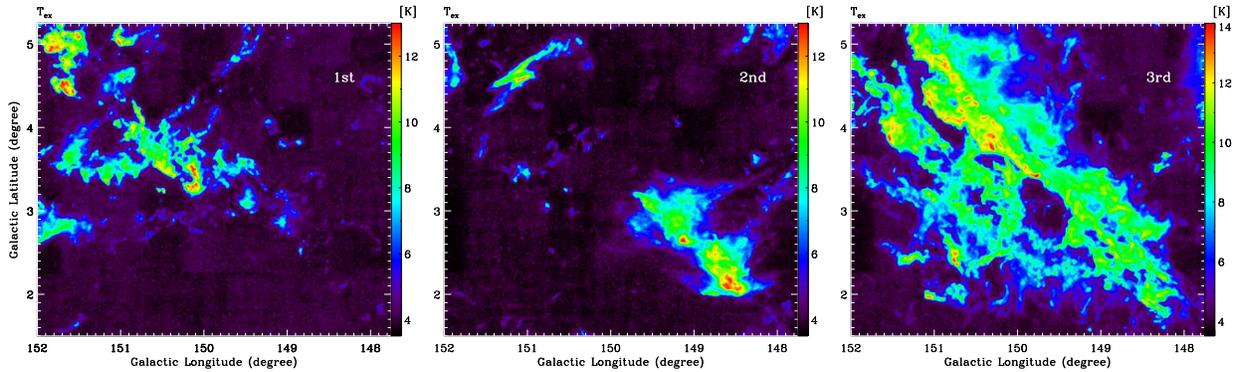}
\caption{Excitation temperature maps of the three velocity components identified in the G150 region.}
\label{fig:fig6}
\end{figure}

Assuming the Local Thermodynamic Equilibrium (LTE), we can calculate the excitation temperature of these three components using
\begin{equation}
T_{\rm ex} = T_0 / \ln \left( 1 + \left( \frac{T_{\rm MB}}{(1-e^{-\tau_{\nu}})T_0} + \frac{1}{e^{T_0/T_{\rm bg}}-1} \right)^{-1} \right).
\end{equation}
Here, we use the radiation temperature of $^{12}$CO to calculate the excitation temperature of clouds. In the equation, $T_0 = h\nu/k_{\rm B}$ is the intrinsic temperature of $^{12}$CO, where $h$ is the Planck constant and $k_{\rm B}$ is the Boltzmann constant. $T_{\rm MB}$ is the main beam temperature and $T_{\rm bg}$ is the background temperature with the value of 2.7~K. For $^{12}$CO, the opacity depth $\tau_{\nu} \gg 1$, which means $1-e^{-\tau_{\nu}} \approx 1$. Under all these conditions, we derive the excitation temperature maps and present them in Figure~\ref{fig:fig6}. The former two components have similar excitation temperatures with the same mean value of $\sim$6.2~K, while the mean excitation temperature of the third one is $\sim$7.4~K.

\begin{figure}
\epsscale{1.0}
\plotone{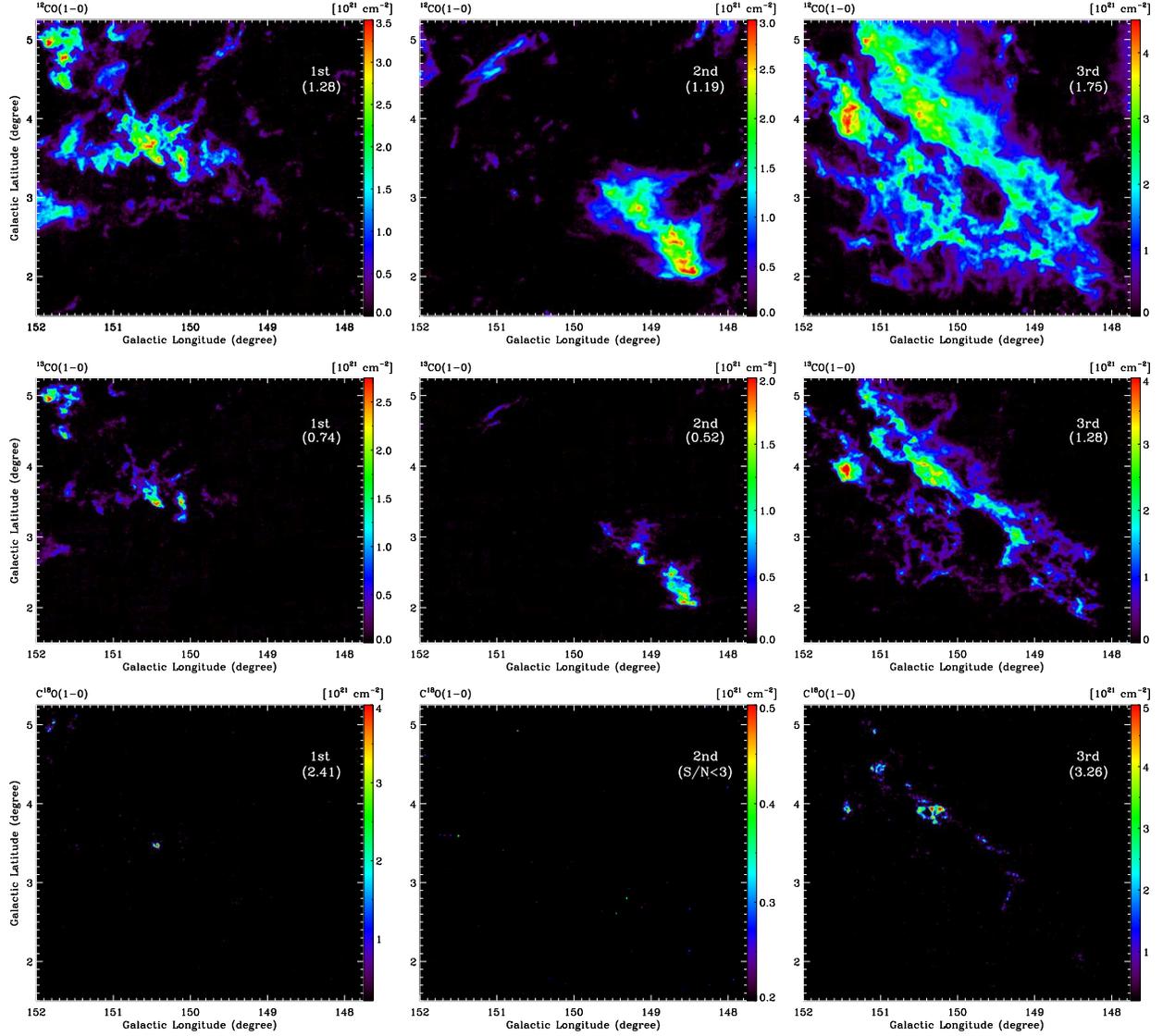}
\caption{H$_2$ column density maps of the three velocity components traced by the $^{12}$CO emission (top panels), $^{13}$CO emission (middle panels), and C$^{18}$O emission (bottom panels). Numbers in the parentheses are the mean values of corresponding column density, with the unit of 10$^{21}$~cm$^{-2}$.}
\label{fig:fig7}
\end{figure}

Figure~\ref{fig:fig7} shows the H$_2$ column density maps of these three components traced by $^{12}$CO, $^{13}$CO, and C$^{18}$O from top to bottom. For $^{12}$CO, the H$_2$ column density can be directly derived from its integrated intensity by using the $X$ factor (CO-to-H$_2$ conversion factor),
\begin{equation}
N_{\rm H_2} = X \int T_{\rm MB, ^{12}CO} dV,
\end{equation}
where the value of $X$ is 1.8$\times$10$^{20}$~cm$^{-2}$~K$^{-1}$~km$^{-1}$~s \citep{dam01}. The calculation of $^{13}$CO column density can be described as
\begin{equation}
N_{\rm ^{13}CO} = 2.42 \times 10^{14} \cdot \frac{\int T_{\rm MB, ^{13}CO} dV}{1-e^{-T_{\rm 0, ^{13}CO}/T_{\rm ex}}},
\end{equation}
where $T_{\rm 0, ^{13}CO} = h\nu_{\rm ^{13}CO}/k_{\rm B}$ is the intrinsic temperature of $^{13}$CO and $T_{\rm ex}$ is the excitation temperature calculated from $^{12}$CO. To derive the H$_2$ column density traced by $^{13}$CO, we multiply the $^{13}$CO column density by the abundance $N_{\rm H_2}/N_{\rm ^{13}CO}$ with the value of 7$\times$10$^{5}$ \citep{fre82}. The method to derive the H$_2$ column density traced by C$^{18}$O is almost the same, the formula of C$^{18}$O column density is
\begin{equation}
N_{\rm C^{18}O} = 2.24 \times 10^{14} \cdot \frac{\int T_{\rm MB, C^{18}O} dV}{1-e^{-T_{\rm 0, C^{18}O}/T_{\rm ex}}},
\end{equation}
and the abundance $N_{\rm H_2}/N_{\rm C^{18}O}$ is 7$\times$10$^{6}$ \citep{cas95}. We notice that the C$^{18}$O emission of second component has a signal-to-noise ratio smaller than three, which means we have not detected the effective signals of C$^{18}$O emission in the second component. So the H$_2$ column density traced by C$^{18}$O of the second component is not available. With the integrated intensity of CO emission as the weight, we also calculate the mean values of H$_2$ column density, which are marked on each panel in Figure~\ref{fig:fig7}. The third component has much greater column density than the other two components.

In the following sections, we mainly focus on the third velocity component ($-$0.5 -- 6.5~km~s$^{-1}$), which is a filamentary molecular cloud and has greater intensity and column density of the $^{13}$CO and C$^{18}$O emission, and we study the properties of this filamentary cloud.

\subsection{Filamentary Molecular Cloud}\label{sec:identify}
\begin{figure}
\epsscale{1.0}
\plotone{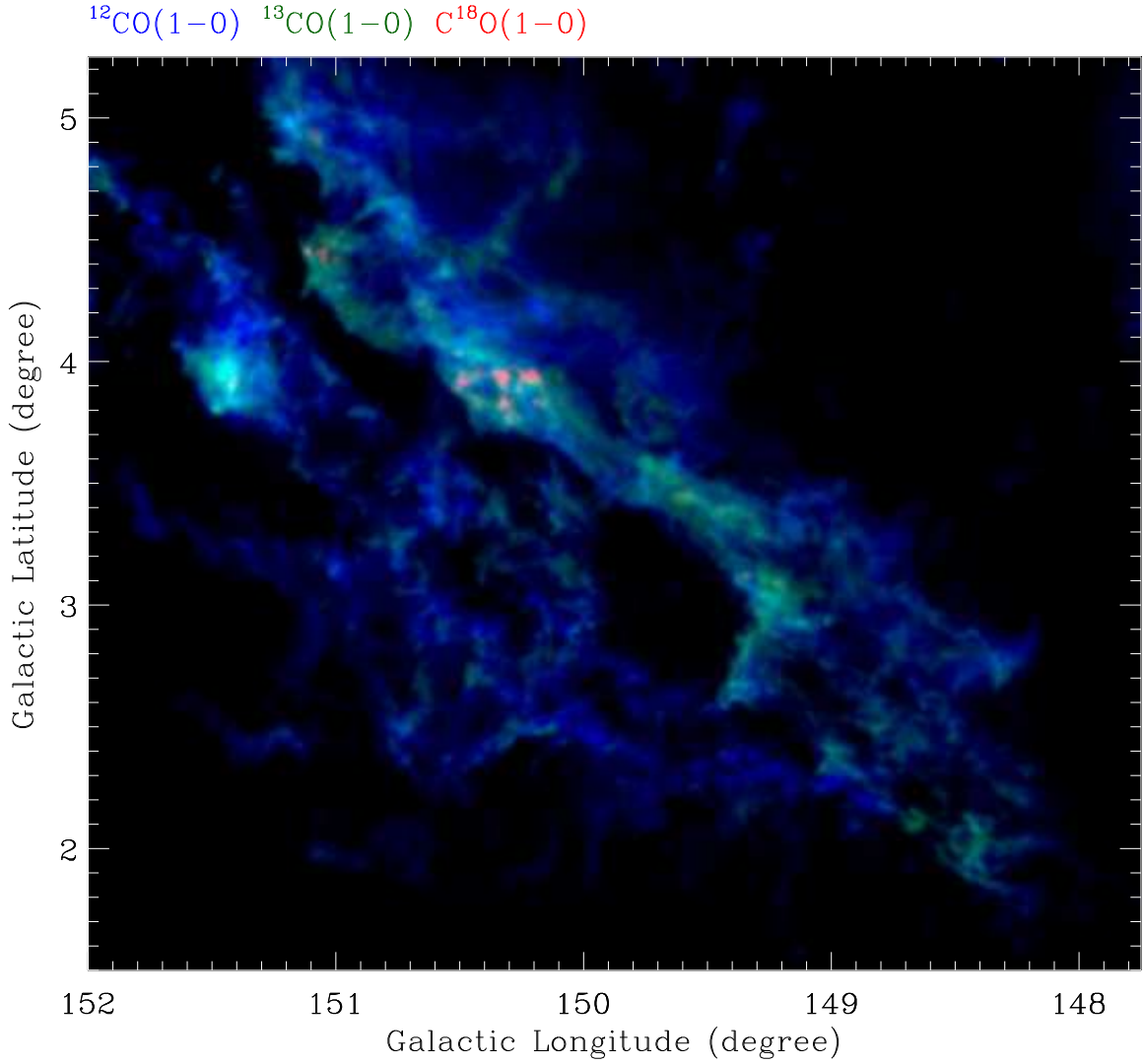}
\caption{Three-color image of the filamentary molecular cloud. The blue area represents the integrated intensity of the $^{12}$CO emission with a velocity range of $-$0.5 to 6.5~km~s$^{-1}$, the green area represents the integrated intensity of the $^{13}$CO emission with a velocity range of 0 to 6.5~km~s$^{-1}$, and the red area represents the integrated intensity of the C$^{18}$O emission with a velocity range of 1.3 to 4.3~km~s$^{-1}$.}
\label{fig:fig8}
\end{figure}

Figure~\ref{fig:fig8} shows the three-color image of the third velocity component. Three CO isotopologues present different views of this filamentary molecular cloud. The $^{12}$CO emission (blue area) outlines the general structure of the cloud, the $^{13}$CO emission (green area) reveals the skeleton of the cloud, and the C$^{18}$O emission (red area) only appears at the brightest parts of the $^{13}$CO emission (green area). This is mainly because that $^{12}$CO represents the diffuse gas with the low density of $\sim$10$^{2}$~cm$^{-3}$, while $^{13}$CO traces the denser intermediate gas with the medium density of $\sim$10$^{3}$~cm$^{-3}$ and C$^{18}$O traces the densest inner gas with the high density of $\sim$10$^{4}$~cm$^{-3}$.

According to the distribution of $^{13}$CO emission, we can clearly see that there is more than one filamentary structure within this molecular cloud. In order to identify these filaments, we adopt the visual inspection method consisting of the following steps in this work. First, we check the $^{13}$CO integrated intensity of this molecular cloud and search for the elongated structures with intensities greater than three times the rms noise level. Second, we calculate the length and width of each derived elongated structure to check whether the aspect ratio is larger than three or not, and regard the qualified ones as filament candidates. Third, we inspect the $^{13}$CO channel maps to make sure these candidates are coherent structures distributed in the adjacent channels rather than different velocity components along the line of sight, overlapping with each other to mimic the shape of the filament. After applying these steps, we finally identify 12 filaments, which are presented in Figure~\ref{fig:fig9}, Figure~\ref{fig:fig10}, and Figure~\ref{fig:fig11}. Each filament is named after the first letter of ``filament'' and Arabic numerals from ``1'' to ``12''. The length and width of identified filaments are listed in Table~\ref{tab:tab1}.

As pointed out by \citet{mye09} and \citet{hil11}, one of the most important criteria to make a distinction between main filament and sub-filaments is the difference of column density, which can be calculated from the integrated intensity. Shown in Figure~\ref{fig:fig9}, F1 is the main filament with higher intensity of the $^{13}$CO emission compared to the rest filaments. With lower integrated intensity, F2, F3, F4, and F5 are located on the northeastern side of F1, and F6, F7, F8, and F9 are located on the southwestern side. These eight filaments in the surrounding area of F1 can be considered to be the sub-filaments. In the eastern area, F10, F11, and F12 are distributed approximately parallel to F1. Their integrated intensity is not as high as F1, so we may also consider them to be the sub-filaments.

Figure~\ref{fig:fig10} and Figure~\ref{fig:fig11} present the velocity distributions of the identified filaments. Each filament only appears in the adjacent channels (Figure~\ref{fig:fig10}) and does not have a large velocity gradient in itself (Figure~\ref{fig:fig11}), indicating that it is the self-consistent structure instead of different components overlapping along the line of sight. In Figure~\ref{fig:fig10}, F2, F3, F6, F7, F8, F9, and F1 are mainly distributed in almost the same channels from [1.5, 2.5] to [3.5, 4.5], while F4 and F5 are mainly distributed in [3.5, 4.5] and [4.5, 5.5]. In Figure~\ref{fig:fig11}, F2, F3, F6, F8, and F1 have similar velocity components from 2 to 4~km~s$^{-1}$, while F4 and F5 with the velocity mainly from 3.5 to 5~km~s$^{-1}$ and F7 and F9 with the velocity mainly from 1.5 to 3~km~s$^{-1}$ are slightly different with F1. Considering the spatial distribution of these eight filaments (F2 to F9) and F1 presented in Figure~\ref{fig:fig9}, it is reasonable to believe that they are associated with F1. Away from the main filament, F10 ($\sim$2.5~km~s$^{-1}$), F11 ($\sim$1.5~km~s$^{-1}$), and F12 ($\sim$1.5~km~s$^{-1}$) have different velocities from F1. These three filaments may not be associated with F1.

The spatial organization of the identified filaments is more similar to the ``ridge-nest'' structure \citep{hil11}, rather than the ``hub-filament'' \citep{mye09} and ``filament-striation'' \citep{pal13} structures. The ``hub-filament'' model requires the central body ``hub'' to be of low aspect ratio, while in our case, the filament in the central area is more like a ``ridge'' with high aspect ratio. We also have not found the faint ``striations'' closely perpendicular to the main filament, which are the main features of ``filament-striation'' model. In addition, the main filament in the inner area is uni-directional with higher integrated intensity and the sub-filaments in the outer area are variedly directional with lower integrated intensity, which are the characteristics of ``ridge-nest'' structure.

Figure~\ref{fig:fig12} is the integrated intensity map of the C$^{18}$O emission. With respect to the other two isotopologues, C$^{18}$O traces the densest parts of this filamentary cloud. Only the main filament F1 and sub-filaments F2, F3, and F6 are identified in C$^{18}$O. The C$^{18}$O distributions of these filaments appear to be discontinuous along their length. In addition, we notice that there is an isolated clump with strong emission to the east of F1, which also appears at the brightest part of the $^{12}$CO and $^{13}$CO emission.

\subsection{Properties of Filaments}\label{sec:properties}
We present the CO spectra of all the identified filaments in Figure~\ref{fig:fig13}. Each filament's spectra are averaged within the area surrounding it where the intensity of the $^{13}$CO emission is three times higher than the rms noise level. The $^{13}$CO emission is stronger in F1, F2, and F6, and decrease to below 2~K in other sub-filaments. In F10, F11, and F12, which are the sub-filaments away from F1, the $^{13}$CO emission is only about 1~K. The emission of C$^{18}$O is clearly detected in F1 and F2. For F3 and F6, of which we have detected the C$^{18}$O emission in the integrated intensity map (Figure~\ref{fig:fig12}), the C$^{18}$O emission in their spectra is too weak to be identified. This is mainly due to the fact that the area of emission is too small compared to the area where spectra are averaged. For other sub-filaments, the C$^{18}$O emission is unidentifiable.

\begin{figure}
\epsscale{1.0}
\plotone{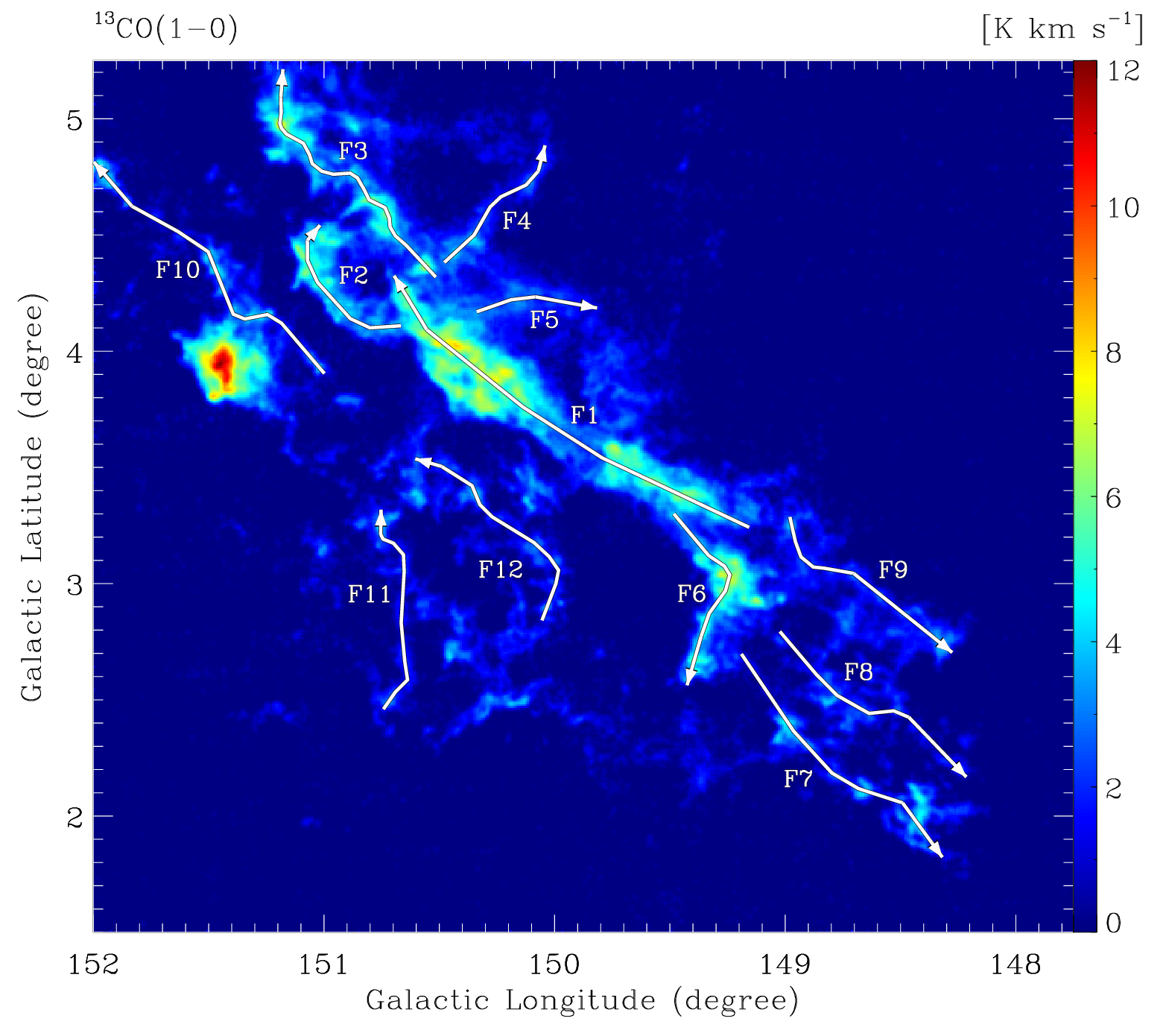}
\caption{Integrated intensity map of filamentary structures in the $^{13}$CO emission. The velocity range is from 0 to 6.5~km~s$^{-1}$. The solid white lines show the positions of identified filaments, and the arrows indicate the directions of position-velocity plots shown in Figure~\ref{fig:fig14}. The name of each filament is marked beside it.}
\label{fig:fig9}
\end{figure}

\begin{figure}
\epsscale{1.0}
\plotone{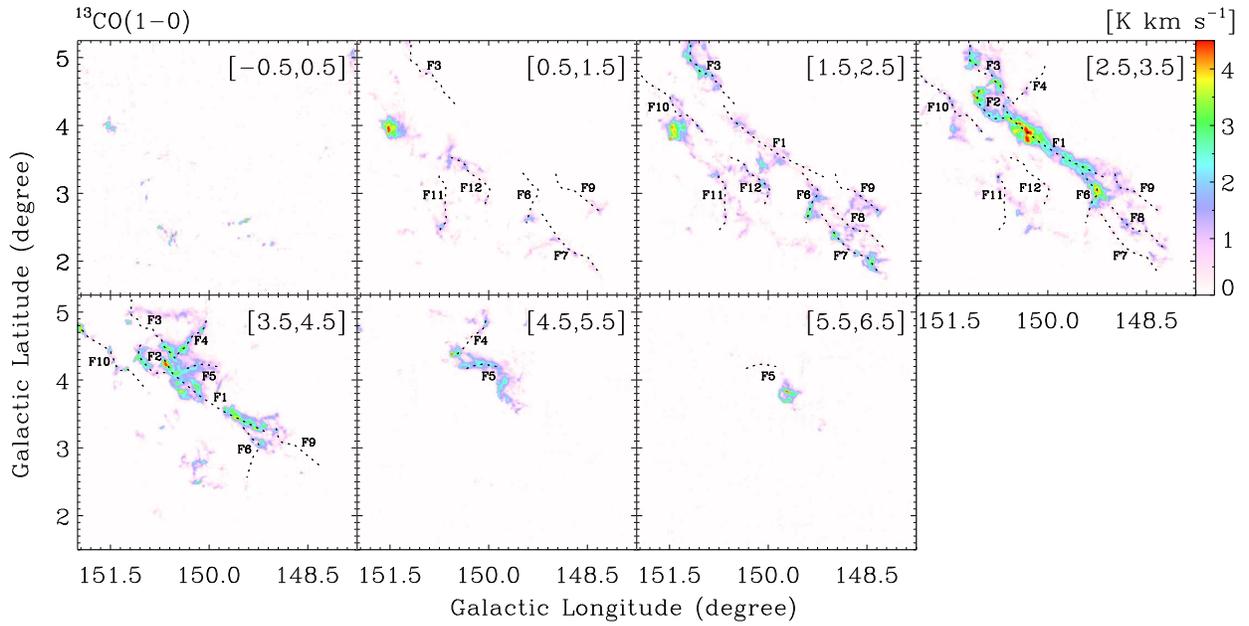}
\caption{Velocity channel maps of filamentary structures in the $^{13}$CO emission. The dashed black lines show the positions of identified filaments. The name of each filament is marked beside it. The velocity range (in the unit of km~s$^{-1}$ ) marked on each map is the integration range of each channel.}
\label{fig:fig10}
\end{figure}

\begin{figure}
\epsscale{1.0}
\plotone{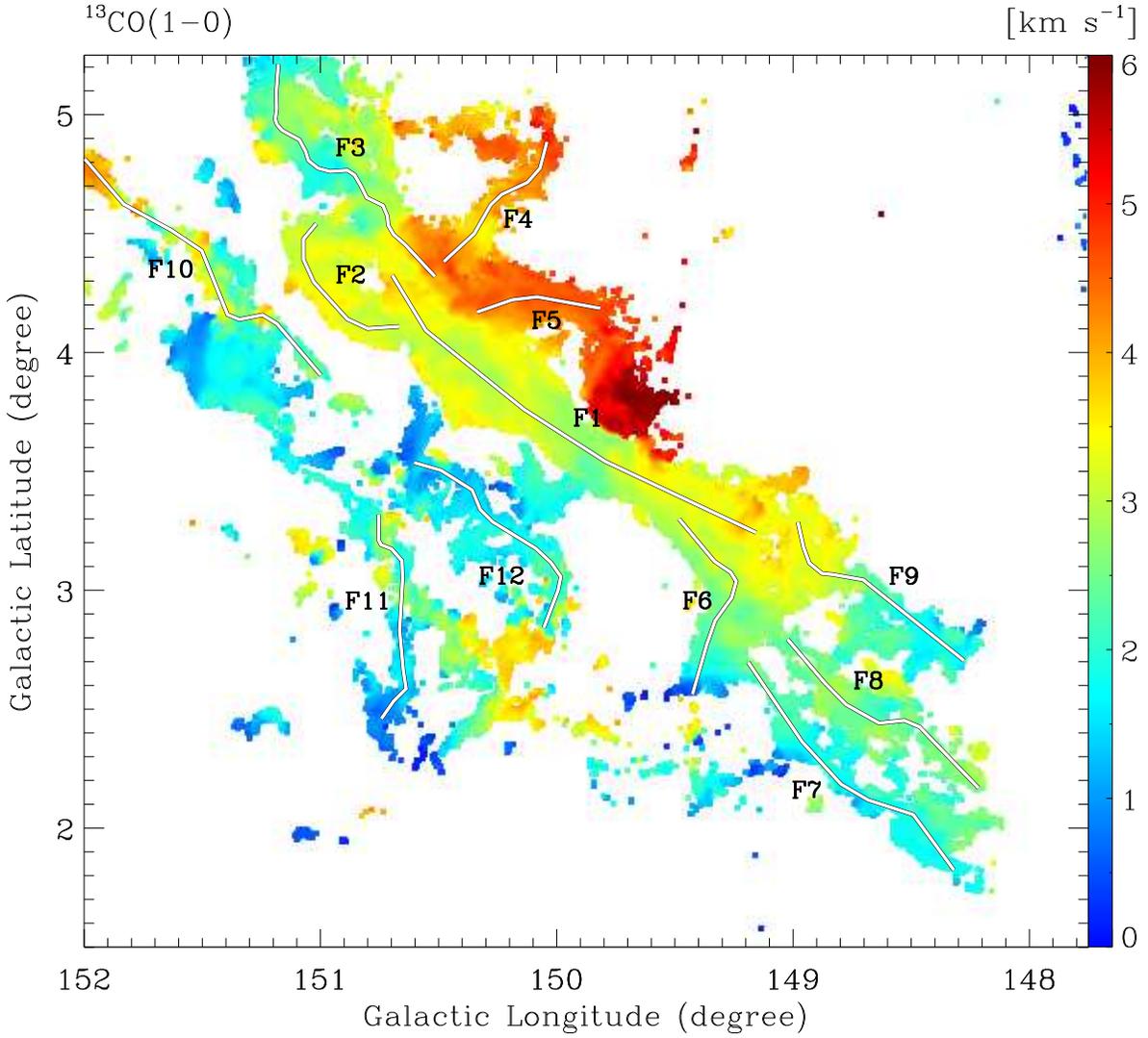}
\caption{Velocity distribution map of filamentary structures in the $^{13}$CO emission. The solid white lines show the positions of identified filaments. The name of each filament is marked beside it.}
\label{fig:fig11}
\end{figure}

\begin{figure}
\epsscale{1.0}
\plotone{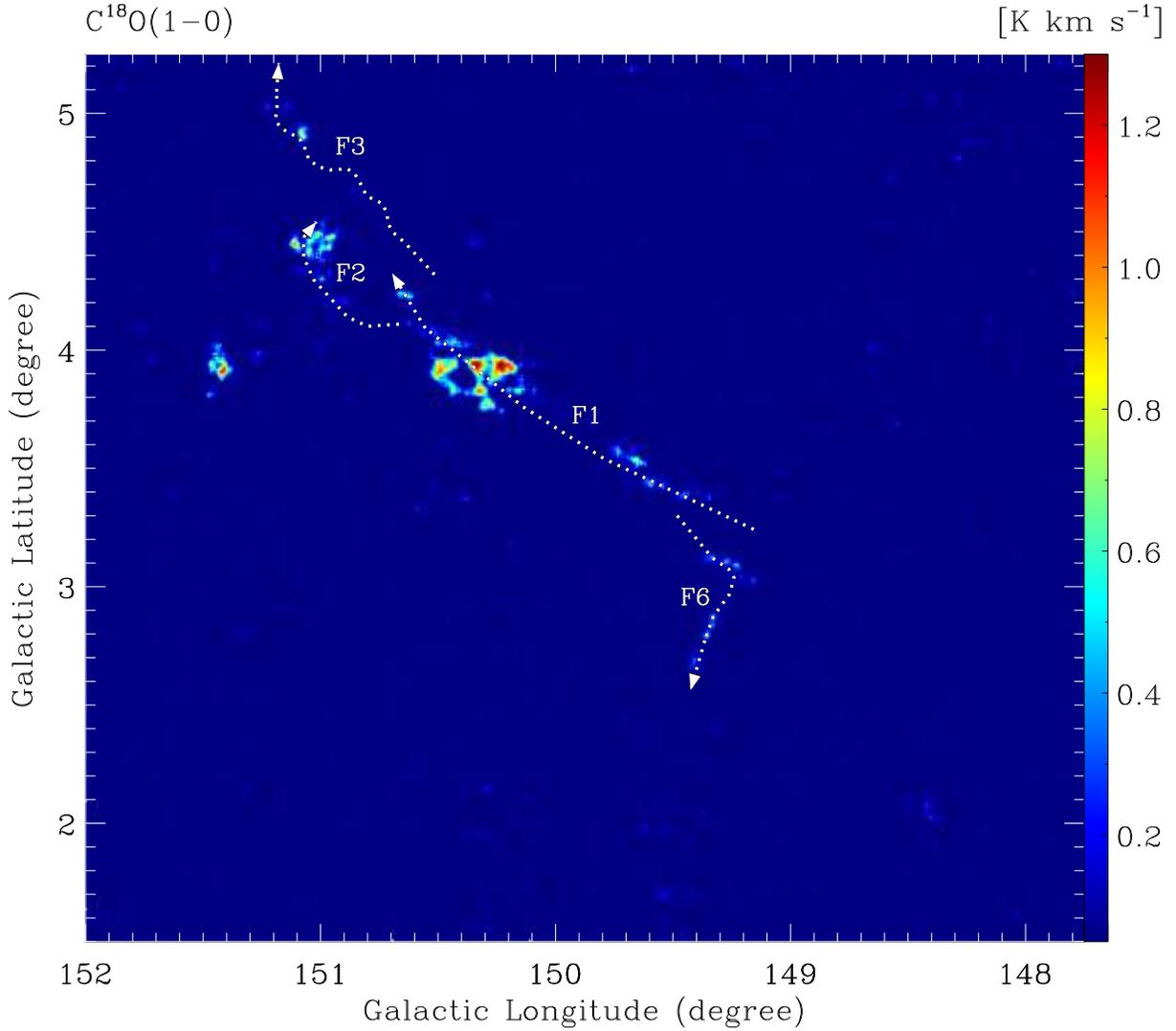}
\caption{Integrated intensity map of filamentary structures in the C$^{18}$O emission. The velocity range is from 1.3 to 4.3~km~s$^{-1}$. The dotted white lines show the positions of identified filaments, and the arrows indicate the directions of position-velocity plots showed in Figure~\ref{fig:fig15}. The name of each filament is marked beside it.}
\label{fig:fig12}
\end{figure}

\begin{figure}
\epsscale{1.0}
\plotone{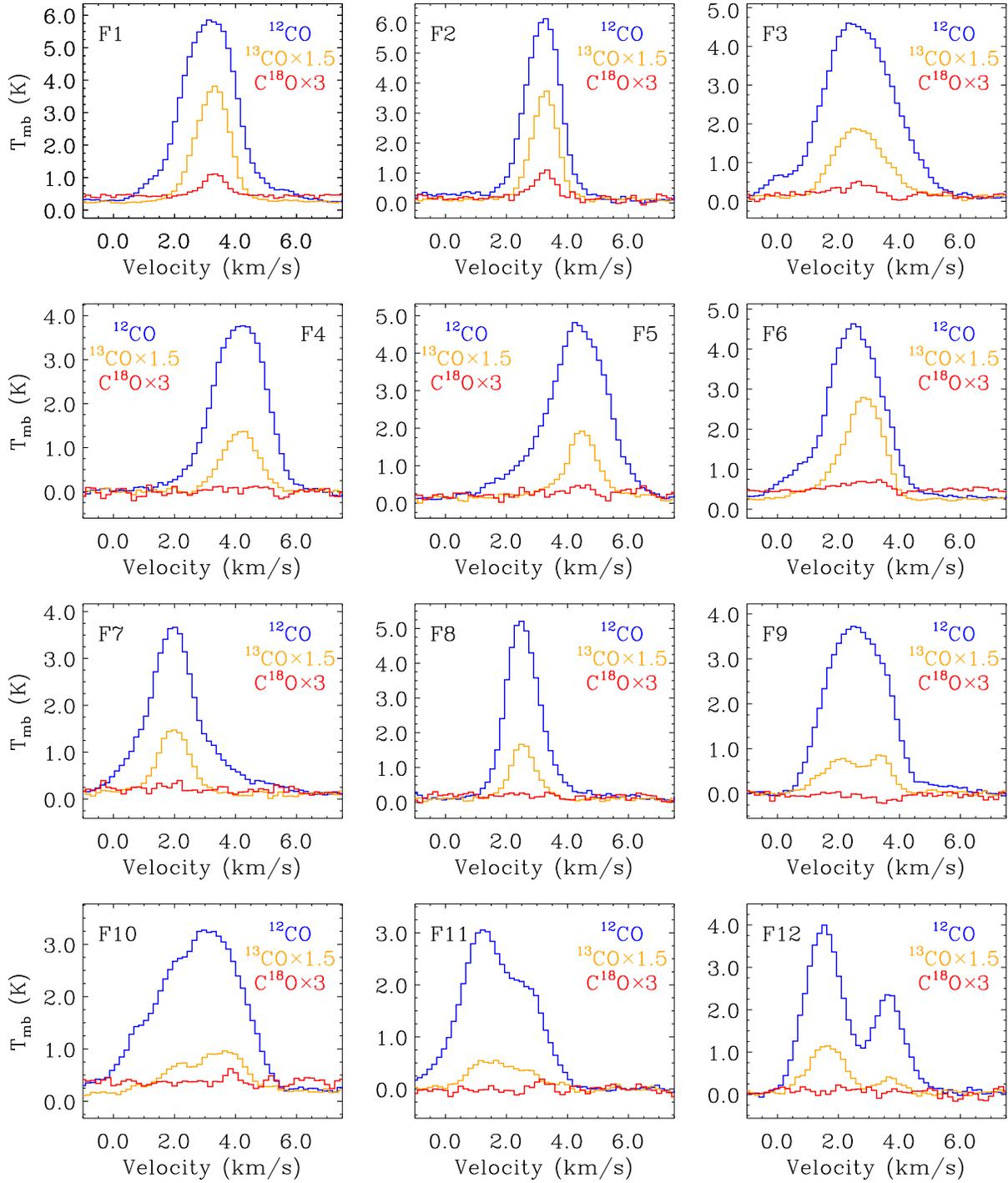}
\caption{CO spectra of the identified filaments. The blue spectrum shows the $^{12}$CO emission, the orange spectrum shows the $^{13}$CO emission multiplied by a factor of 1.5, and the red spectrum shows the C$^{18}$O emission multiplied by a factor of three.}
\label{fig:fig13}
\end{figure}

\begin{figure}
\epsscale{0.9}
\plotone{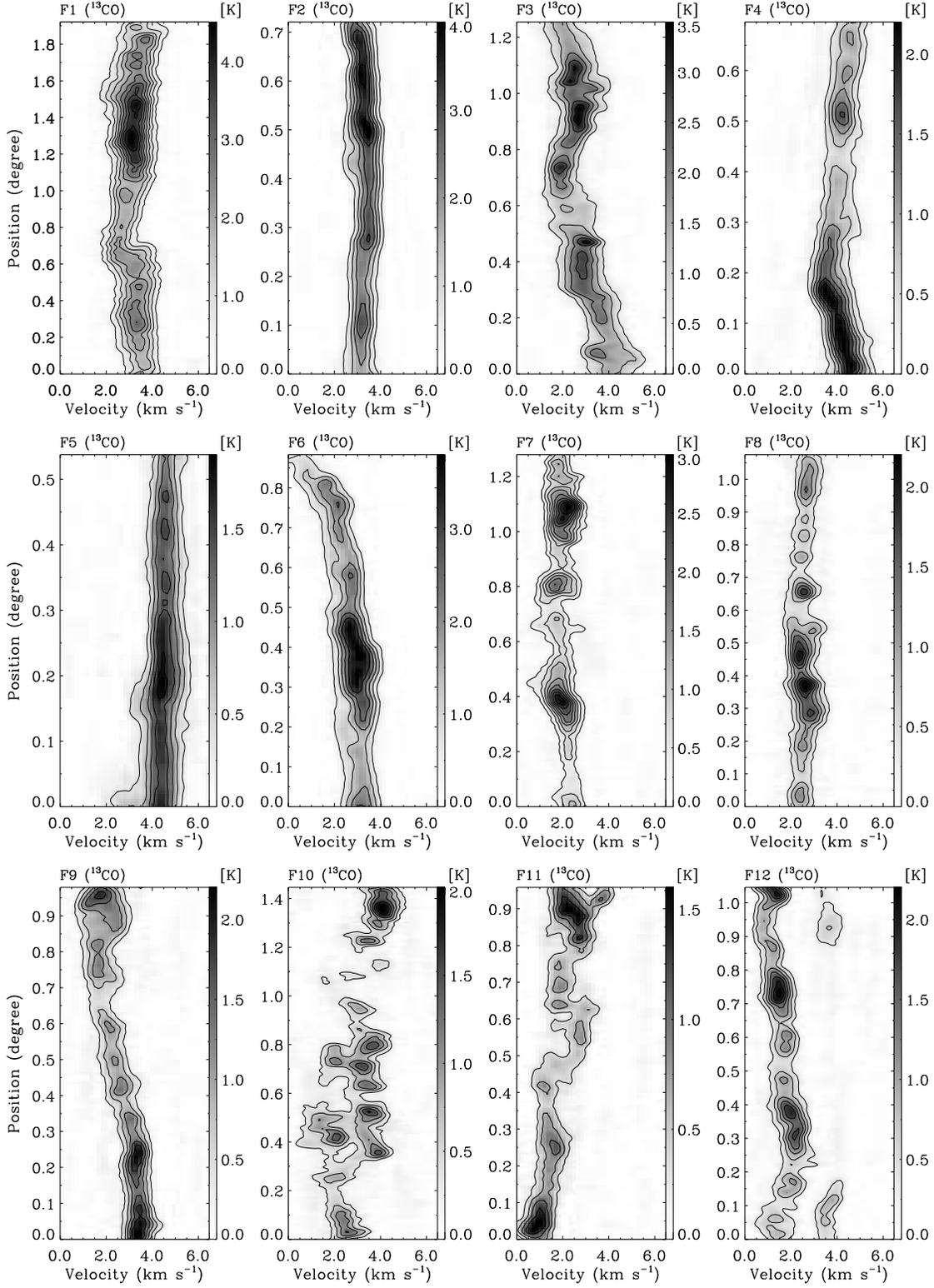}
\caption{$^{13}$CO position-velocity plots of the identified filaments extracted along the solid arrowed lines shown in Figure~\ref{fig:fig9} with widths of 15$\arcmin$ for F1, 10$\arcmin$ for F2, F3, and F6, and 8$\arcmin$ for the rest. The contours are overlaid from 10~$\sigma$ with an interval of 10~$\sigma$ for F1, F2, F3, and F6, and from 5~$\sigma$ with an interval of 5~$\sigma$ for the rest (``$\sigma$'' is the rms noise level in each plot).}
\label{fig:fig14}
\end{figure}

In Figure~\ref{fig:fig14}, we show the position-velocity plots of the filaments in the $^{13}$CO emission. These position-velocity plots are extracted along the solid white lines marked in Figure~\ref{fig:fig9} with the directions indicated by the white arrows. The main filament and most sub-filaments display continuous structures without significant curvature along the position axes and do not have other evident components along the velocity axes. The plot of F3 presents a twisted velocity structure with a velocity gradient from southeast (4~km~s$^{-1}$) to northwest (2~km~s$^{-1}$) (referring to the directions of filaments in Figure~\ref{fig:fig9}) of it. F6 also has a velocity gradient, which is from 3 to 0.5~km~s$^{-1}$ occurring at its southern part. Different from other filaments, F9, F10, and F11 all have high $^{13}$CO emission parts located at their two ends, with velocity gradients from northwest (3.5~km~s$^{-1}$) to southeast (1.5~km~s$^{-1}$) for F9, from southeast (2.5~km~s$^{-1}$) to northwest (4~km~s$^{-1}$) for F10, and from south (1~km~s$^{-1}$) to north (3~km~s$^{-1}$) for F11, making their $^{13}$CO spectra shown in Figure~\ref{fig:fig13} have two peaks. Corresponding to its spectra in Figure~\ref{fig:fig13}, F12 has another component on the velocity axes, but this one is too weak in the $^{13}$CO emission to be identified as a new filament.

\begin{figure}
\epsscale{1.0}
\plotone{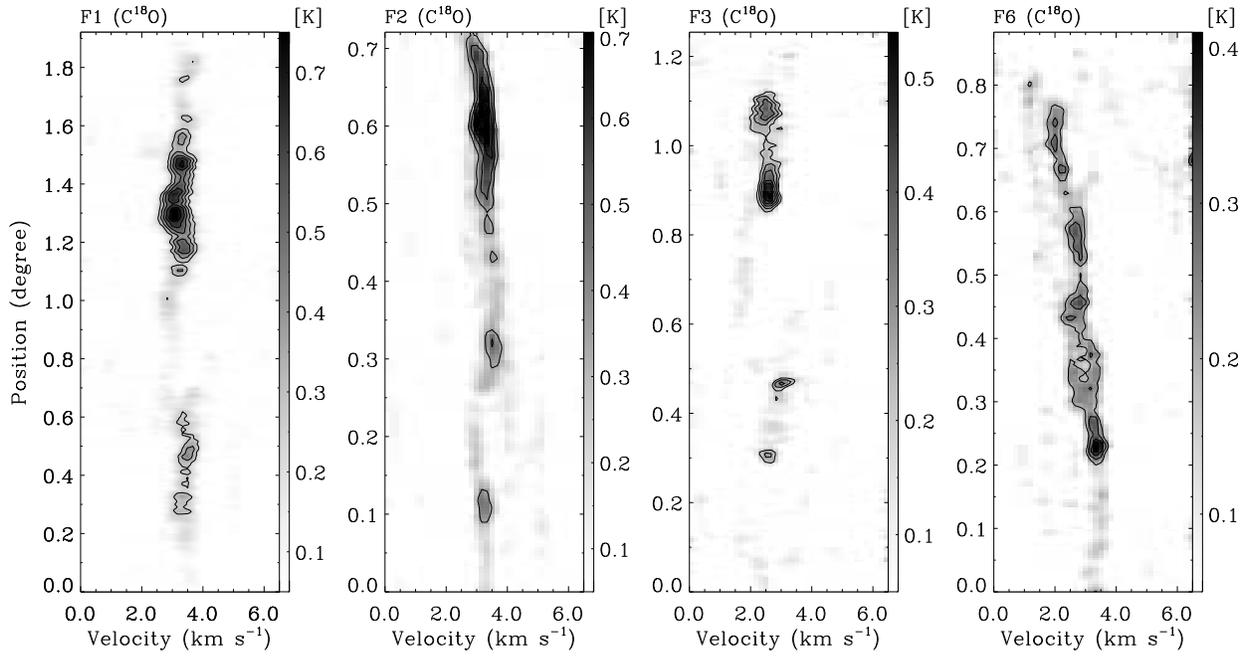}
\caption{C$^{18}$O position-velocity plots of the identified filaments extracted along the dotted arrowed lines shown in Figure~\ref{fig:fig12} with widths of 12.5$\arcmin$ for F1 and 8$\arcmin$ for F2, F3, and F6. The contours are overlaid from 5~$\sigma$ with an interval of 2~$\sigma$ for F1 and F2, and from 3~$\sigma$ with an interval of 1~$\sigma$ for F3 and F6 (``$\sigma$'' is the rms noise level in each plot).}
\label{fig:fig15}
\end{figure}

The C$^{18}$O position-velocity plots of the filaments are presented in Figure~\ref{fig:fig15}. These plots are extracted along the dotted arrowed lines shown in Figure~\ref{fig:fig12}. Only the four filaments detected with the C$^{18}$O emission are presented here. Representing the densest parts of filamentary structures, the C$^{18}$O velocity structures of these four filaments all display discontinuous features along the position axes. Similar with its $^{13}$CO velocity structure in Figure~\ref{fig:fig14}, F6 is identified with the velocity gradient of 3.5 to 2~km~s$^{-1}$ from north to south (referred to the directions of filaments in Figure~\ref{fig:fig12}), while no evident velocity gradient is revealed in the other three filaments.

In Section~\ref{sec:overview}, we have derived the excitation temperature of filamentary structures (right panel in Figure~\ref{fig:fig6}). We notice that F2, F3, and the northern part of F1 have relatively high excitation temperatures ($T_{\rm ex} > \rm 12~K$). The $^{12}$CO gas distribution shows that there is a cleft between F1--F2 and F10--F12, and the main filament F1 has an asymmetric distribution with a diffuse extension at the northern part and sharp cut-off at the cleft. Moreover, the areas with high excitation temperature are located at the edge of the cleft. These morphological features suggest that some external effects, such as UV irradiation, propagation of shock waves, or dynamical interaction between filaments, may be the plausible causes of the enhancement of excitation temperature in F1, F2, and F3. We extract the mean excitation temperature of each filament according to the distribution of filaments in Figure~\ref{fig:fig9}, and list the results in Table~\ref{tab:tab1}.

\begin{figure}
\epsscale{1.0}
\plotone{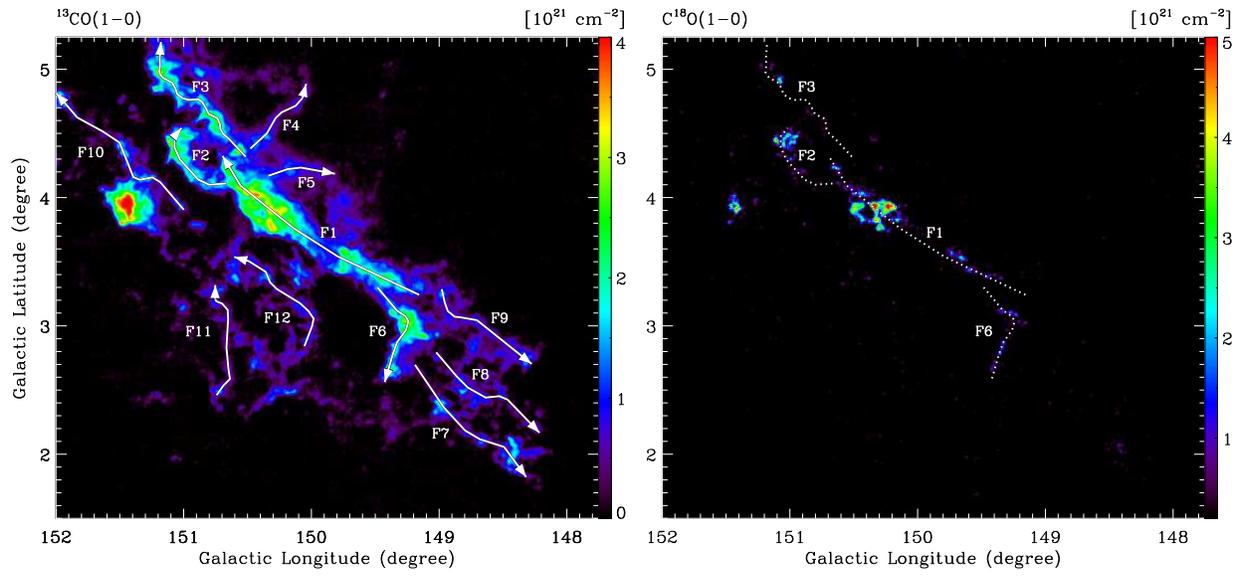}
\caption{H$_2$ column density maps of filamentary structures traced by the $^{13}$CO emission (left panel) and the C$^{18}$O emission (right panel). The solid white lines (left panel) and dotted white lines (right panel) show the positions of identified filaments. The arrows (right panel) indicate the directions used to determine the left and right sides of filaments corresponding to the $R < 0$ and $R > 0$ parts of radial density profiles in Figure~\ref{fig:fig17}. The name of each filament is marked beside it.}
\label{fig:fig16}
\end{figure}

We have also derived the H$_2$ column density maps of filamentary structures (right panels in Figure~\ref{fig:fig7}). Overlaid with the positions of identified filaments, we present the new column density maps in Figure~\ref{fig:fig16}. Similar to the distributions of filaments in the integrated intensity maps (Figure~\ref{fig:fig9} and Figure~\ref{fig:fig12}), the new column density maps in Figure~\ref{fig:fig16} show that the main filament F1 resembles the shape of a ``ridge'' with high column density in the inner area, while the sub-filaments around or away from it form the ``nest'' with lower column densities and various directions, no matter whether they are shown in $^{13}$CO (left panel) or in C$^{18}$O (right panel).

With H$_2$ column density, we can calculate the LTE mass of filaments by
\begin{equation}
M = \mu m_{\rm H} \int N_{\rm H_2} dS,
\end{equation}
where $\mu$ is the mean molecular weight with the value of 2.83, $m_{\rm H}$ is the mass of hydrogen atom, and $S$ is the area of CO emission. We adopt the distance of 400~pc, which will be discussed in Section~\ref{sec:distance}, when calculating the area. We also measure the length of each filament and derive the linear masses, which are listed in Table~\ref{tab:tab1}.

In addition, we calculate the averaged line width of each filament in the following way. First, we calculate the line width of each pixel in the area around each filament where the integrated intensity of the $^{13}$CO emission is greater than three times the rms noise level. Then, we use the integrated intensity of these pixels as the weight to calculate the averaged line width of each area, and thus the line width of each filament. The result is listed in Table~\ref{tab:tab1}. 

\begin{figure}
\epsscale{1.0}
\plotone{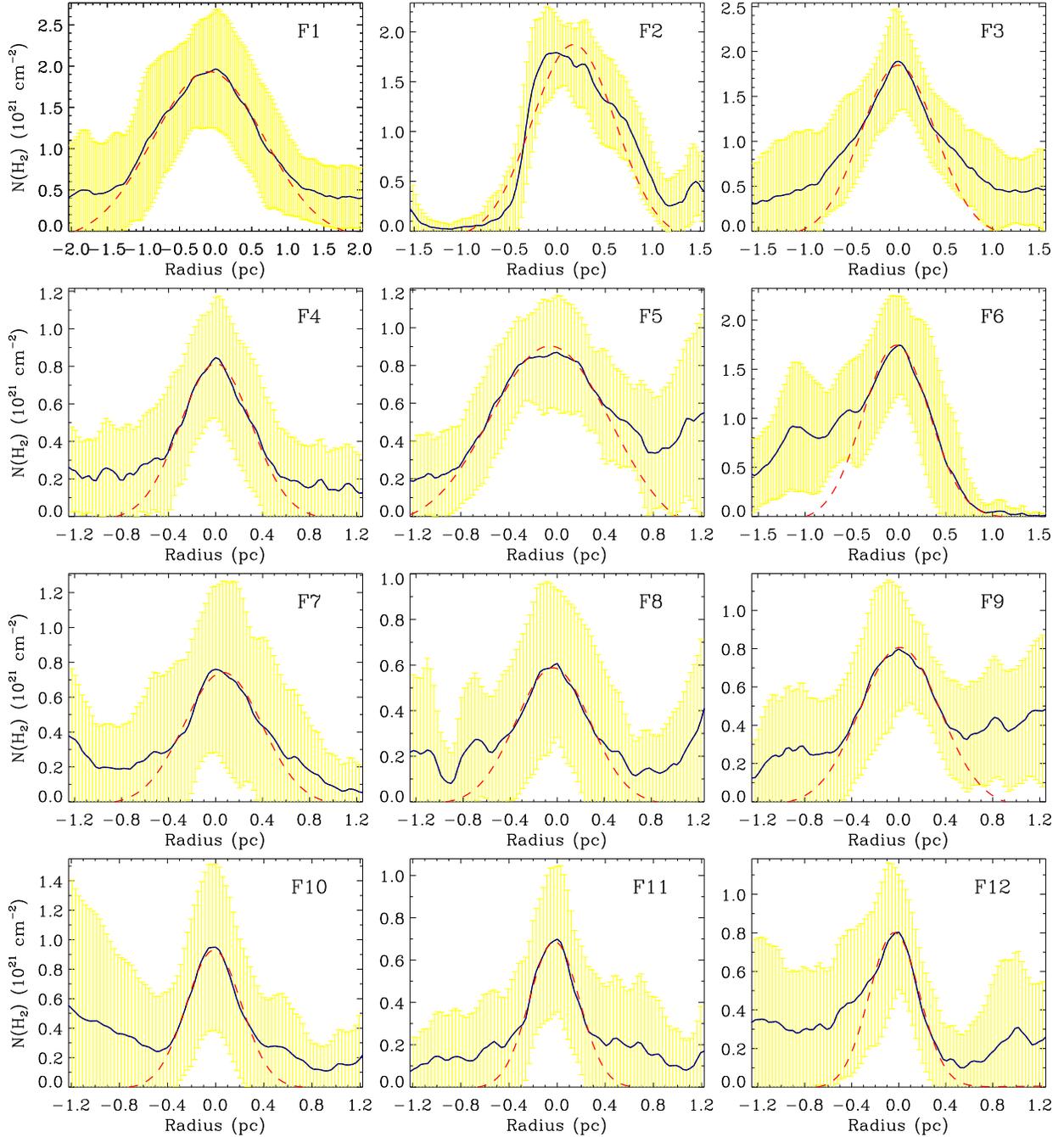}
\caption{Mean radial column density profiles perpendicular to the identified filaments (blue curves). The position of the peak density in each profile is regarded as the center of the profile, thus the position of $R = 0$. The $R < 0$ and $R > 0$ parts of each profile correspond to the left and right sides of each filament in Figure~\ref{fig:fig16}. The yellow areas show the $\pm$~1~$\sigma$ dispersion of the distributions of radial profiles along the filaments. The dashed red curves show the Gaussian fittings to the inner parts of the profiles.}
\label{fig:fig17}
\end{figure}

We further calculate the radial density profiles of filaments based on the H$_2$ column density of filaments traced by $^{13}$CO emission (shown in the left panel of Figure~\ref{fig:fig16}). Similar to \citet{arz11} and \citet{pal13}, we first determine the tangential direction of each pixel along the position of each filament shown in Figure~\ref{fig:fig16}. Then, for each pixel, we derive one column density profile perpendicular to the tangential direction of the pixel. Finally, we average the profiles of all pixels along each filament and derive the mean radial density profile. The results are presented in Figure~\ref{fig:fig17}. The profiles of most filaments have Gaussian-like shapes in the inner parts, and the outer parts of the profiles reflect the distributions of surrounding structures of filaments. We apply Gaussian fittings to the inner parts of the profiles, which are shown by the dashed red curves in Figure~\ref{fig:fig17}.

According to the results of Gaussian fittings, we also calculate the FWHM width of each filament (see Table\ref{tab:tab1}). We note that the widths of our sample of filaments are much larger than those of the filaments identified in the $Herschel$ Gould Belt survey \citep[e.g.,][]{arz11, pal13, cox16}. This is mainly because the far-IR observations of $Herschel$ trace much denser ISMs than the CO gases that are used as tracers in our observations. Another reason is that we choose to fit the inner Gaussian-like portions rather than inner flat portions \citep{arz11, pal13} of the profiles to calculate the widths of the filaments.

Based on the properties listed in Table~\ref{tab:tab1}, we derive the median excitation temperature, line width, length, width, and linear mass of filaments with the values of 9.28~K, 0.85~km s$^{-1}$, 7.30~pc, 0.79~pc, and 17.92~$M_\sun$~pc$^{-1}$, respectively.

\section{DISCUSSION}\label{sec:discussion}

\subsection{Distance of Filaments}\label{sec:distance}
Distance is an essential parameter when calculating the area of emission and the length of filaments. In our observations, we have derived the Local Standard of Rest (LSR) velocity of the molecular gas. However, as mentioned by \citet{xu06}, the kinematic distance calculated from LSR velocity is not accurate because of the large peculiar motions of gas material in the spiral arms of the Milky Way. Another method is to use the trigonometric parallax of the maser to estimate the distance, which is adopted by the Bar and Spiral Structure Legacy (BeSSeL) Survey \citep{rei14} to measure the distances of high-mass SFRs in the Milky Way. However, according to their catalog, they have not derived the distance of any maser source in the G150 region.

\begin{figure}
\epsscale{1.0}
\plotone{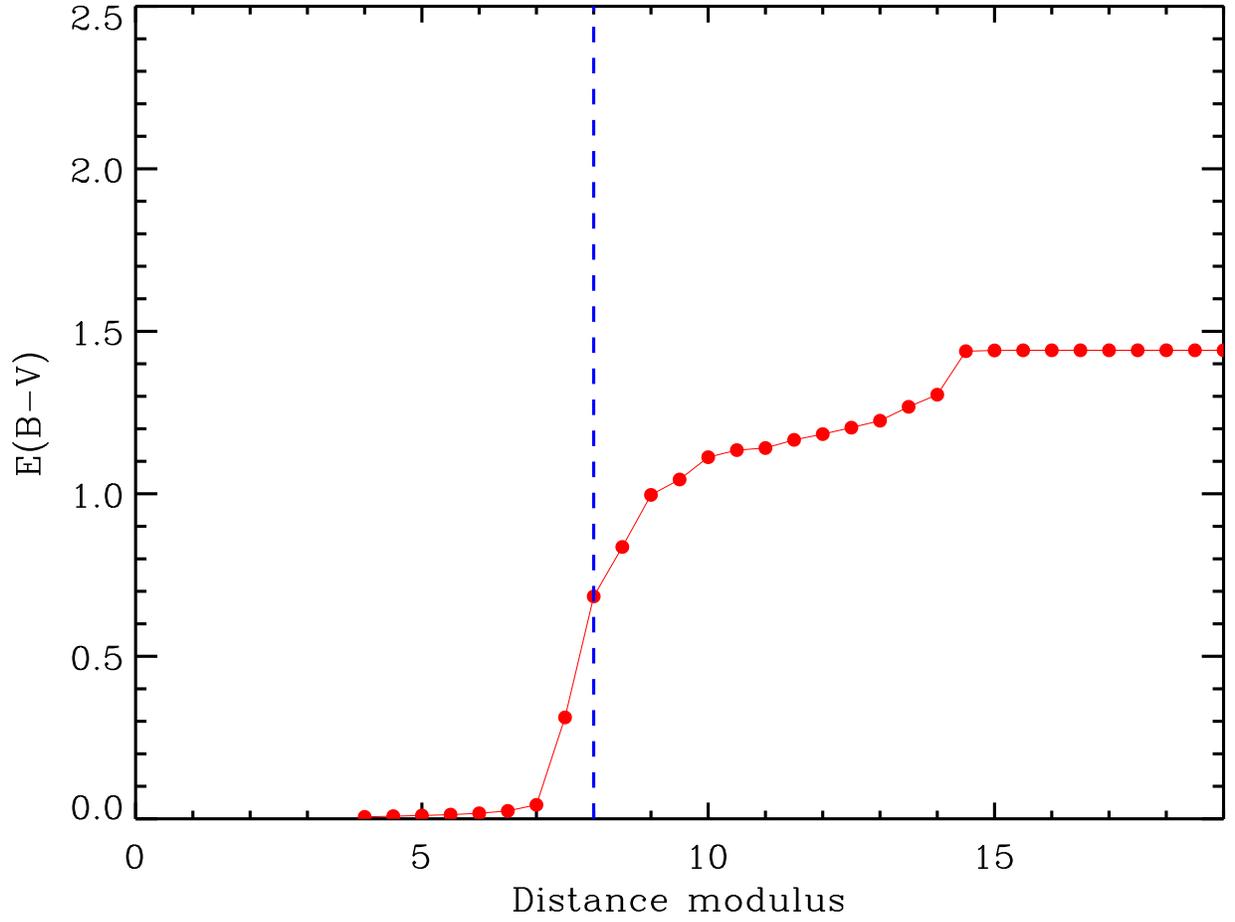}
\caption{Median cumulative reddening in each distance modulus bin of all the selected regions (see Section~\ref{sec:distance}). The dashed blue line marks the distance of our studied filamentary molecular cloud.}
\label{fig:fig18}
\end{figure}

In this work, the distance to the identified filaments is estimated based on the 3D extinction map from \citet{gre15}. With 5-band $grizy$ Pan-STARRS~1 photometry and 3-band 2MASS $JHK_{\rm s}$ photometry of stars embedded in the dust, they trace the extinction on 7$\arcmin$ scales out of a distance of several kiloparsecs, by simultaneously inferring stellar distance, stellar type, and dust reddening along the line of sight. We select six regions with high intensity integrated from $-$0.5 to 6.5~km~s$^{-1}$ (see Figure~\ref{fig:fig9}). The regions are centered at the Galactic coordinates of (150.5, 4.0), (150.3, 3.9), (149.7, 3.5), (151.4, 3.9), (151.1, 4.4), and (149.2, 3.0), respectively, and with the same radius of 0.25$\arcdeg$. In Figure~\ref{fig:fig18}, we show the median cumulative reddening in each distance modulus (DM) bin of all the selected regions from \citet{gre15}. We notice one rapid increase in the extinction centered at DM$\sim$8 (400~pc), which could be due to the dust reddening in the filamentary molecular cloud. Therefore, we take 400~pc as the distance of the filamentary molecular cloud.

\subsection{Comparison with DisPerSE Algorithm}
The filaments presented in this work are all identified by the visual inspection method (see Section~\ref{sec:identify}). To testify our identification, we present the result from the Discrete Persistent Structure Extractor (DisPerSE) algorithm to make a comparison. DisPerSE is a coherent multi-scale identification approach to all kinds of astrophysical structures, especially the filamentary structures, in the large-scale matter distribution of the universe \citep{sou11I}. It is originated from the analysis of the cosmic web and its filamentary network, and has been upgraded to apply to 3D simulated data and X-ray data observed by the satellite $Suzaku$ \citep{sou11II}. As illustrated by \citet{sou11I}, implementation of this method is based on two theories, the Morse theory, which makes it possible for the application of topological principles to astrophysical data, and the persistence theory, which helps to deal with the intrinsic uncertainty and Poisson noise in the data set.

\begin{figure}
\epsscale{1.0}
\plotone{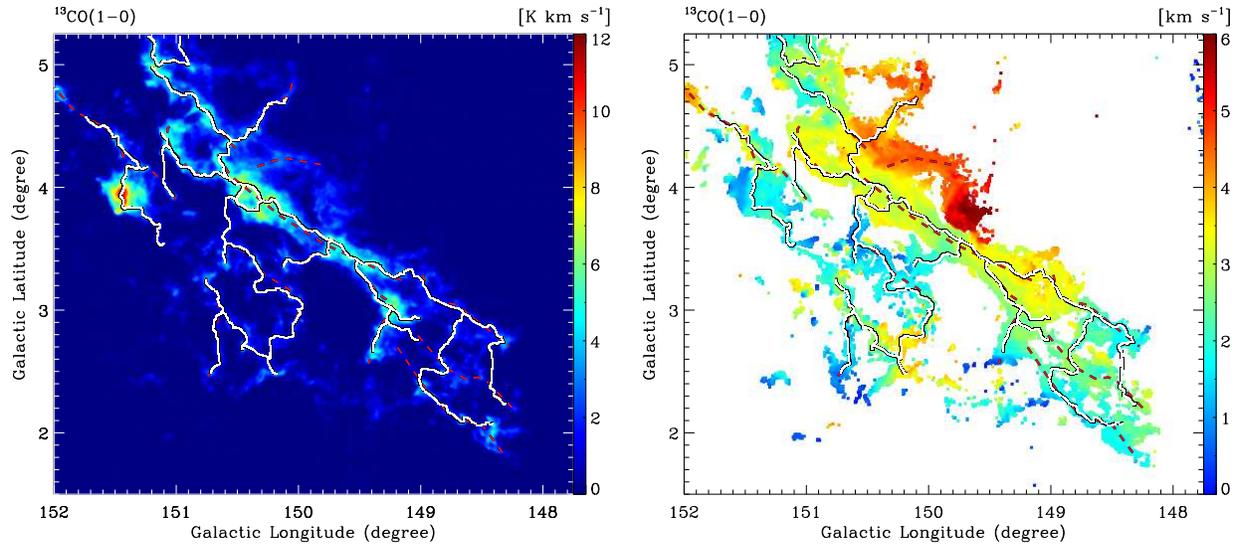}
\caption{Filamentary structures identified by the DisPerSE algorithm overlaid on the integrated intensity map (left panel) and velocity distribution map (right panel) of the $^{13}$CO emission. The solid white lines indicate the positions of filaments DisPerSE identified and the dashed red lines indicate the positions of filaments identified by visual inspection.}
\label{fig:fig19}
\end{figure}

In the practical process, we first define a persistence level with the value of three times the rms noise level, and the algorithm works over the Delaunay tessellation of the $^{13}$CO data to obtain a discrete density field by the Delaunay Tessellation Field Estimator technique \citep{sch00, van09}. Then the algorithm computes discrete Morse complexes from the density field and extracts the filamentary structures. The result is shown in Figure~\ref{fig:fig19}. The solid white lines mark the filaments identified by DisPerSE, with the backgrounds of the integrated intensity map of the $^{13}$CO emission (left panel) and the velocity distribution map of the $^{13}$CO emission (right panel), and the dashed red lines are the filaments identified by visual inspection.

In the left panel of Figure~\ref{fig:fig19}, the result of DisPerSE is quite consistent with our result, except for the filaments F5 and F8. DisPerSE tends to extract filaments as a set of connected segments \citep{sou11II}, while we prefer to consider the continuity of filaments in spatial distribution during the identification. In the right panel, the connections between some segments by DiePerSE seem too random, ignoring the velocity gradients between the segments, while the filaments we identified are consistent with the velocity distribution (presented in Section~\ref{sec:identify}). We argue that DisPerSE can track the trails of most filamentary structures, but we have not found that it can take the velocity information into account when connecting the segments to form the shape of the filament.

\subsection{Gravitational Stability of Filaments}
Theoretical studies of self-gravitating cylinders predict that cylinders should have a maximum, critical linear mass above which cylinders will radially collapse into a line \citep{jac10}. Same as the cylinders, the critical linear masses of filaments determine their gravitational stability. If the mass of a filament is greater than the critical value, the filament is gravitationally unstable and will fragment into clumps along its length, which are expected to evolve into protostars. Indeed, young stellar objects (YSOs) are observed to be distributed along the supercritical filaments in, e.g., Taurus \citep{gol08} and Aquila \citep{bon10}. On the contrary, with mass lower than the critical value, the filament is gravitationally unbound and maybe in an expanding state, and is even expected to disperse during the turbulent crossing time, which is $\sim$0.3~Myr for a typical subcritical filament \citep{arz13}, unless confined by the external pressure \citep{fis12}.

The critical linear mass can be described as $M_{\rm crit} = 2c_{\rm s}^{2}/G$ \citep{ost64} when the thermal pressure dominates over the turbulent pressure. In this formula, $c_{\rm s}$ is the isothermal sound speed, which is $\sim$0.2~km~s$^{-1}$ when the gas temperature is 10~K \citep{arz13}, and $G$ is the gravitational constant, which is given as 1/232~km$^2$~s$^{-2}$~$M_{\sun}^{-1}$~pc \citep{sol87}. Since the mean excitation temperature of our identified filaments is around 10~K, we can estimate a critical linear mass of $M_{\rm crit}$ $\approx$ 18.56~$M_{\sun}$~pc$^{-1}$. In the case of turbulence dominance, we use a virial linear mass of $M_{\rm vir} = 2\sigma_{\rm tot}^{2}/G$ \citep{fie00} as the critical linear mass, where $\sigma_{\rm tot}$ is the total velocity dispersion, which can be calculated as
\begin{equation}
\sigma_{\rm tot} = \sqrt{\frac{k_{\rm B} T_{\rm kin}}{\mu m_{\rm H}} + \sigma_{\rm NT}^{2}} = \sqrt{\frac{k_{\rm B} T_{\rm kin}}{\mu m_{\rm H}} + \sigma_{\rm obs}^{2} - \sigma_{\rm T,obs}^{2}}.
\end{equation}
In the calculation, $\mu$ is the mean weight of molecular gas with a value of 2.83, and $\sigma_{\rm NT}$ is the non-thermal velocity dispersion, which can be obtained by subtracting the thermal velocity dispersion ($\sigma_{\rm T,obs}$) from the observed velocity dispersion ($\sigma_{\rm obs}$). We have derived the $^{13}$CO line width ($\Delta v$) of filaments, so the observed velocity dispersion ($\sigma_{\rm obs}$) is $\Delta v/\sqrt{8 \rm ln 2}$. Furthermore, the thermal portion of the velocity dispersion is $\sigma_{\rm T,obs} = \sqrt{\frac{k_{\rm B} T_{\rm kin}}{\mu_{\rm obs} m_{\rm H}}}$, where $T_{\rm kin}$ is the kinetic temperature that equals to the excitation temperature and $\mu_{\rm obs}$ is the molecular weight of the observed molecule, which is 29 for $^{13}$CO.

\begin{figure}
\epsscale{1.0}
\plotone{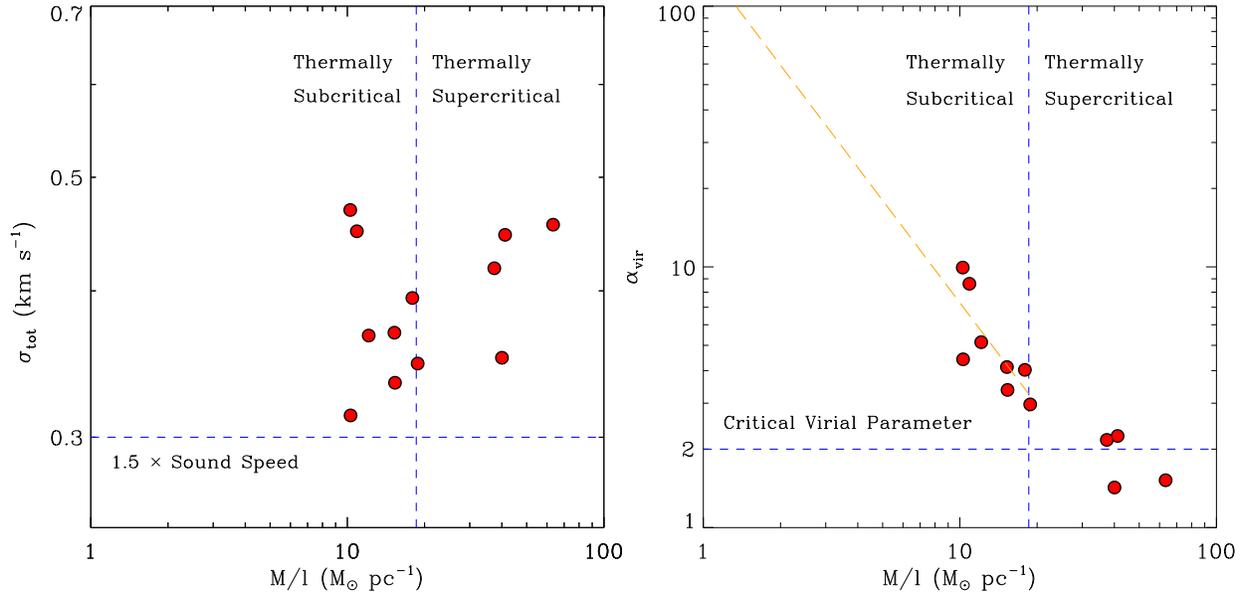}
\caption{Left panel: the relationship between total velocity dispersion and LTE linear mass for the identified filaments. The vertical dashed blue line is the critical linear mass $M_{\rm crit}$ $\approx$ 18.56~$M_{\sun}$~pc$^{-1}$. The horizontal dashed blue line is 1.5 times of the isothermal sound speed, thus $\sim$0.3~km~s$^{-1}$. Right panel: the relationship between the virial parameter and LTE linear mass for the identified filaments. The vertical dashed blue line is the critical linear mass $M_{\rm crit}$ $\approx$ 18.56~$M_{\sun}$~pc$^{-1}$. The horizontal dashed blue line is the critical virial parameter with the value of two. The dashed orange line is power-law fitting result of the correlation with an index of $-$1.30~$\pm$~0.5.}
\label{fig:fig20}
\end{figure}

The left panel of Figure~\ref{fig:fig20} shows the relationship between total velocity dispersion ($\sigma_{\rm tot}$) and LTE linear mass ($M/l$) of our identified filaments. Only four filaments (F1, F2, F3, and F6) are identified as thermally supercritical filaments with their linear masses significantly greater than the critical linear mass ($M_{\rm crit}$ $\approx$ 18.56~$M_{\sun}$~pc$^{-1}$), while most of the rest filaments are thermally subcritical. We notice that linear mass of F5 (17.92~$M_{\sun}$~pc$^{-1}$) and linear mass of F9 (18.80~$M_{\sun}$~pc$^{-1}$) are close to the critical mass. Considering the errors of estimated distance mentioned in Section~\ref{sec:distance}, it is hard to identify whether these two filaments are thermally supercritical or thermally subcritical. The four thermally supercritical filaments are exactly the ones detected with the C$^{18}$O emission (see Section~\ref{sec:identify}), indicating that they are much denser than the other filaments and more likely to be supercritical. All of the filaments, no matter  whether they are thermally subcritical or thermally supercritical, have velocity dispersions of 1.5 times higher than the isothermal sound speed ($c_{\rm s}$ $\approx$ 0.2~km~s$^{-1}$), indicating that the turbulent motions may play an important role in the stability of filaments. Both subcritical and supercritical filaments do not have clear relationships between velocity dispersion and linear mass, the mean velocity dispersion of the subcritical is $\sim$0.38~km~s$^{-1}$, while the value is $\sim$0.42~km~s$^{-1}$ for the supercritical.

\citet{arz13} suggested that thermally supercritical filaments tend to be virialized and gravitationally bound and, conversely, thermally subcritical filaments are not virialized or gravitationally unbound. The relationship between the virial parameter $\alpha_{\rm vir} = M_{\rm vir}/(M/l)$ and the linear mass ($M/l$) of the filaments is shown in the right panel of Figure~\ref{fig:fig20}. All the thermally subcritical filaments are far from virialized state with high virial parameters ($\alpha_{\rm vir} \gg 2$), and thus tend to be gravitationally unbound. While the supercritical filaments F1 and F2 have virial parameters smaller than two, indicating that they are virialized and gravitationally bound, and F3 and F6 have virial parameters slightly lager than 2, which are 2.24 and 2.16, respectively, close to virialized. Our result is quite similar to the result in \citet{arz13}. The relationship between the virial parameter and the linear mass of subcritical filaments can be fitted as $\alpha_{\rm vir} \propto (M/l)^{-1.30 \pm 0.5}$, of which the index is comparable to the one ($-$0.95~$\pm$~0.12) in \citet{arz13} under the certainty.

Thus, in our sample, all of the thermally subcritical filaments are gravitationally unbound, and thus could be in an expanding state or a stable state when they are confined by external pressure. For the thermally supercritical filaments, two of them are gravitationally bound and expected to fragment into clumps that would evolve into protostars in the future, while the other two filaments are close to being virialized. The large turbulent motions ($\sigma_{\rm tot} \gg c_{\rm s}$) found in our sample of filaments may also support the view that filaments are formed by turbulent compression of interstellar gas \citep{pad01, arz13}.

\subsection{Clumps and YSOs along Filaments}\label{sec:clumps&ysos}
\begin{figure}
\epsscale{1.0}
\plotone{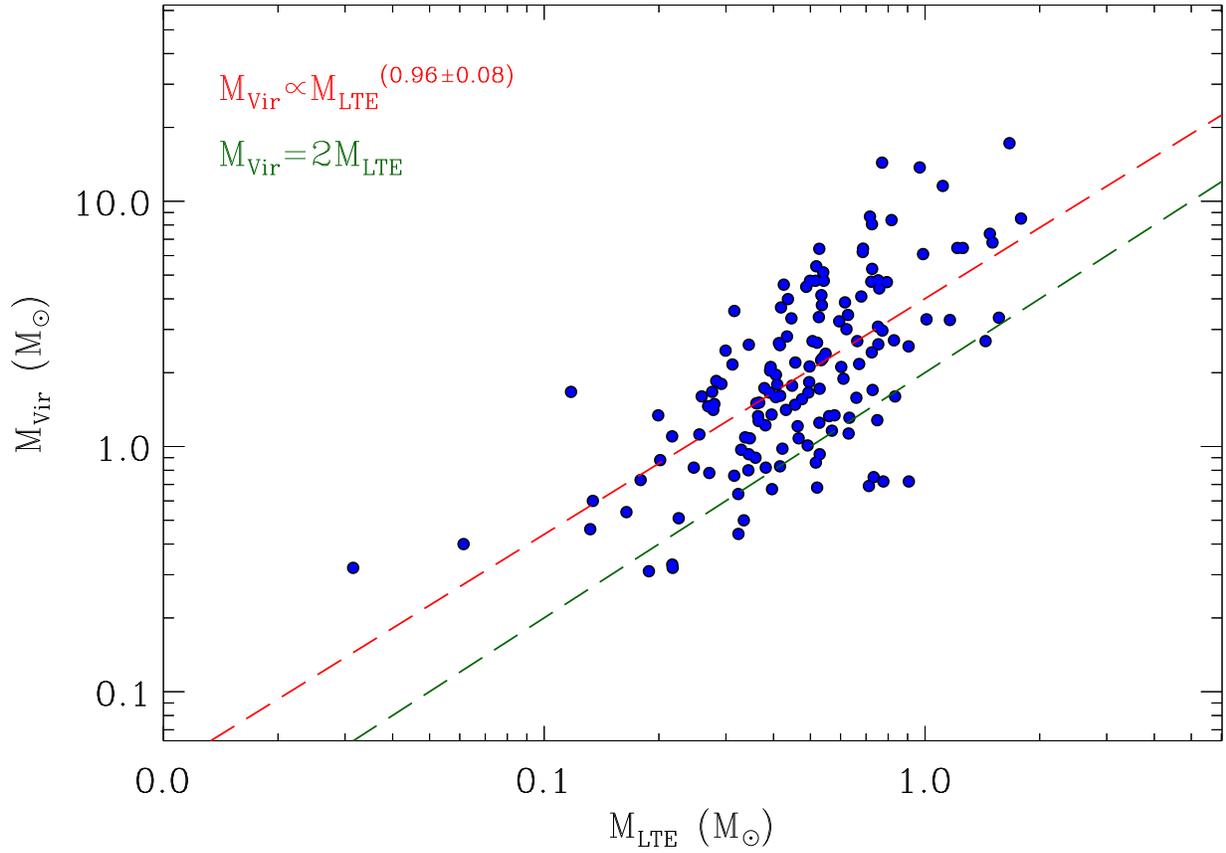}
\caption{Relationship between virial mass and LTE mass of identified clumps. The dashed red line is the power-law fitting result of the correlation with an index of 0.96~$\pm$~0.08. The dashed green line indicates the positions where virial parameter equals two.}
\label{fig:fig21}
\end{figure}

We use the GaussClumps algorithm \citep{stu90} in the CUPID package of Starlink software to identify the clumps within the filamentary molecular cloud. The data we use is $^{13}$CO FITS cubes with a velocity range from $-$0.5 to 6.5~km~s$^{-1}$. We set the parameter Thresh, which determines the minimum peak amplitude of clumps fitted by the algorithm, as three times the rms noise level and other parameters as the default values. The morphologies of the clumps are checked by visual inspection after running the GaussClumps algorithm. As a result, 146 clumps in $^{13}$CO data are identified. We measure the radius, line width, excitation temperature, LTE mass, and virial mass of each clump, which are listed in Table~\ref{tab:tab2}, with median values of 7.26$\times$10$^{-2}$~pc, 0.35~km~s$^{-1}$, 10.04~K, 0.49~$M_{\sun}$, and 1.80~$M_{\sun}$, respectively. We compare the virial mass ($M_{\rm Vir}$) and LTE mass ($M_{\rm LTE}$) of the clumps in Figure~\ref{fig:fig21} and find that about 18 clumps ($\sim$12$\%$) are under virial equilibrium with their virial parameters smaller than two. The mass relationship can be fitted with a power law of $M_{\rm Vir} \propto M_{\rm LTE}^{0.96 \pm 0.08}$. This power-law index we derived is slightly larger than the index 0.75 obtained in CO clumps of North American and Pelican Nebula by \citet{zha14}, and close to the index 0.97 obtained in CO clumps of Gemini cloud by \citet{li15}.

\begin{figure}
\epsscale{1.0}
\plotone{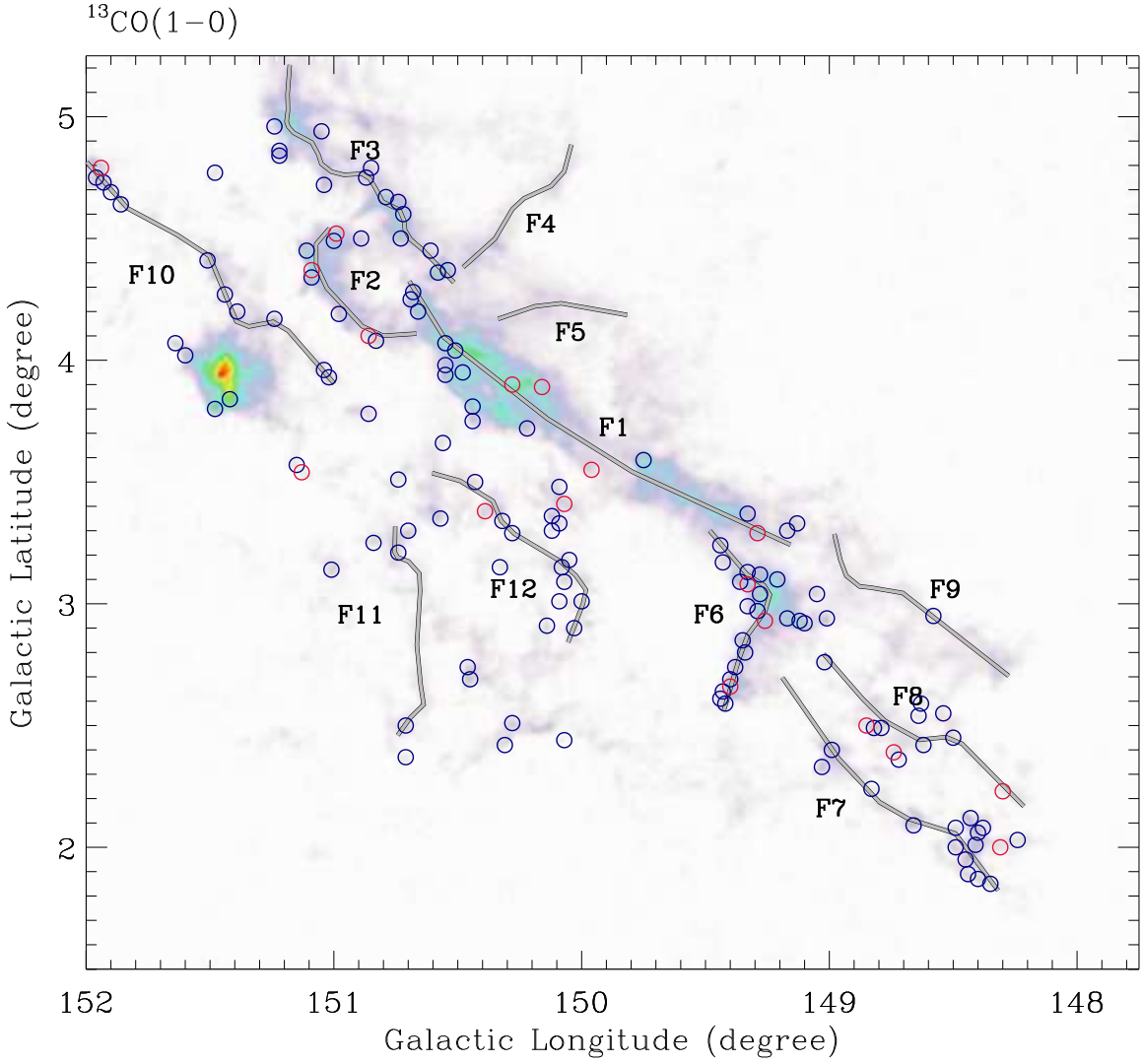}
\caption{Distribution of identified clumps overlaid on the integrated intensity map of the $^{13}$CO emission. The blue and red circles show the positions of unvirialized and virialized clumps, respectively. The solid white lines show the positions of identified filaments. The name of each filament is marked beside it.}
\label{fig:fig22}
\end{figure}

\citet{and10} and \citet{men10} pointed out that there is a close correspondence between the spatial distribution of dense cores and the filamentary network. In our sample of clumps and filaments, we also find this kind of correspondence. Figure~\ref{fig:fig22} illustrates the positions of identified clumps. In the 146 CO clumps, 113 clumps ($\sim$77$\%$) are located within the area around the filaments, where the integrated intensity of the $^{13}$CO emission is greater than three times the rms noise level. A similar result was presented in \citet{hen10}, they found that about 75$\%$ of the detected pre- and protostellar cores are located within the filamentary IRDC G011.11$-$0.12. \citet{kon15} also found that about 71$\%$ -- 78$\%$ of prestellar cores lie within a 0.1~pc width \citep{arz11} of filaments' footprints traced by the DisPerSE algorithm in the Aquila cloud. They also noticed that prestellar cores are preferentially found in the thermally supercritical filaments with a percentage of 66$\%$ -- 75$\%$. In our sample, the percentage is $\sim$44$\%$ for the clumps associated with the thermally supercritical filaments (F1, F2, F3, and F6). For the virialized CO clumps, 10 clumps (56$\%$) are found to be associated with supercritical filaments F1, F2, and F6, and 7 clumps (40$\%$) are associated with virialized filaments (F1 and F2). We also notice that there is a small group of virialized clumps located in subcritical filaments F7 and F8.

\begin{figure}
\epsscale{1.0}
\plotone{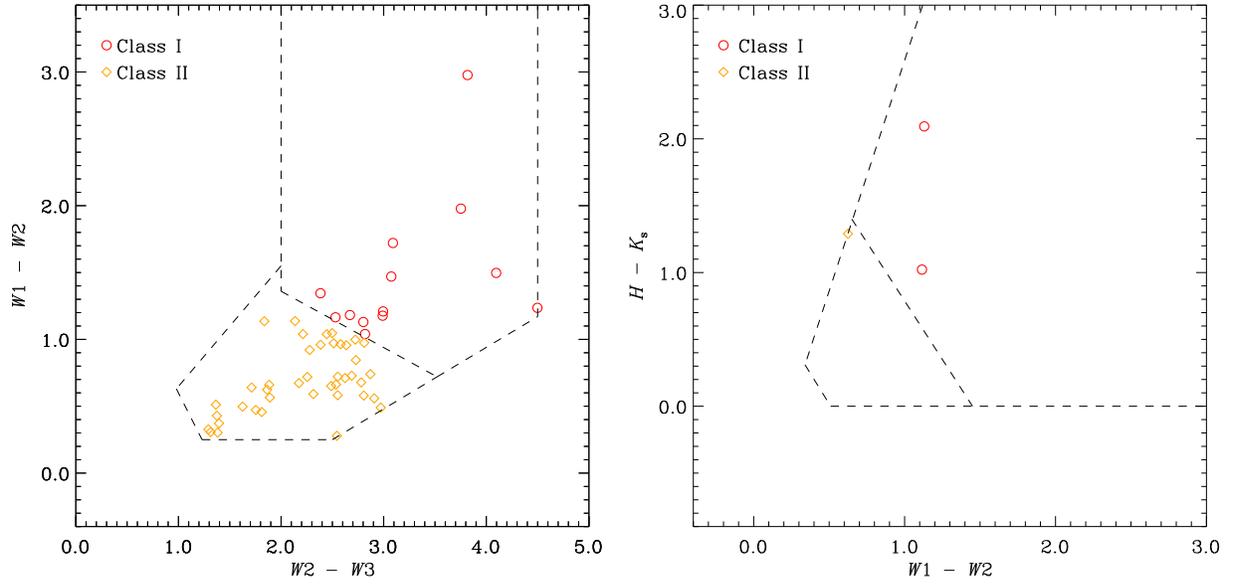}
\caption{Classification of YSO candidates in the WISE $W1 - W2$ versus $W2 - W3$ color-color diagram (left panel) and 2MASS $H - K_{\rm s}$ versus WISE $W1 - W2$ color-color diagram (right panel). The red circles indicate the Class~I objects and the orange diamonds indicate the Class~II objects. The dashed lines represent the classification criteria used in \citet{koe14}.}
\label{fig:fig23}
\end{figure}

With the IR data from 2MASS and WISE surveys, we also investigate the YSO candidates in the area of filamentary structures. YSOs can be classified into disk-bearing YSOs and diskless YSOs according to whether the circumstellar disks exit or not. The IR emission excess created by dusty circumstellar disks makes the IR colors of disk-bearing YSOs different from that of diskless YSOs. On the other hand, the diskless YSOs are unable to be identified only based on their IR colors. So, in this work, we only investigate the disk-bearing YSOs, namely Class~I and Class~II objects, according to the YSOs identification and classification scheme provided by \citet{koe14}. Following the scheme, we first remove the star-forming galaxies and broad-line active galactic nuclei (AGNs) as extragalactic contaminants according to their locations in the WISE $W1 - W2$ versus $W2 - W3$ and $W1$ versus $W1 - W3$ color-color diagrams \citep[see the detailed description in][]{koe14}. Then we select the YSO candidates based on their locations in the WISE $W1 - W2$ versus $W2 - W3$ color-color diagram, which is shown in the left panel of Figure~\ref{fig:fig23}. With the combination of 2MASS $H$ and $K_{\rm s}$ bands, we use $H - K_{\rm s}$ versus $W1 - W2$ color-color diagram, shown in the right panel of Figure~\ref{fig:fig23}, to search for additional YSO candidates among previously unclassified objects. The IR photometric magnitudes and classification of all the identified YSO candidates are listed in Table~\ref{tab:tab3}.

\begin{figure}
\epsscale{1.0}
\plotone{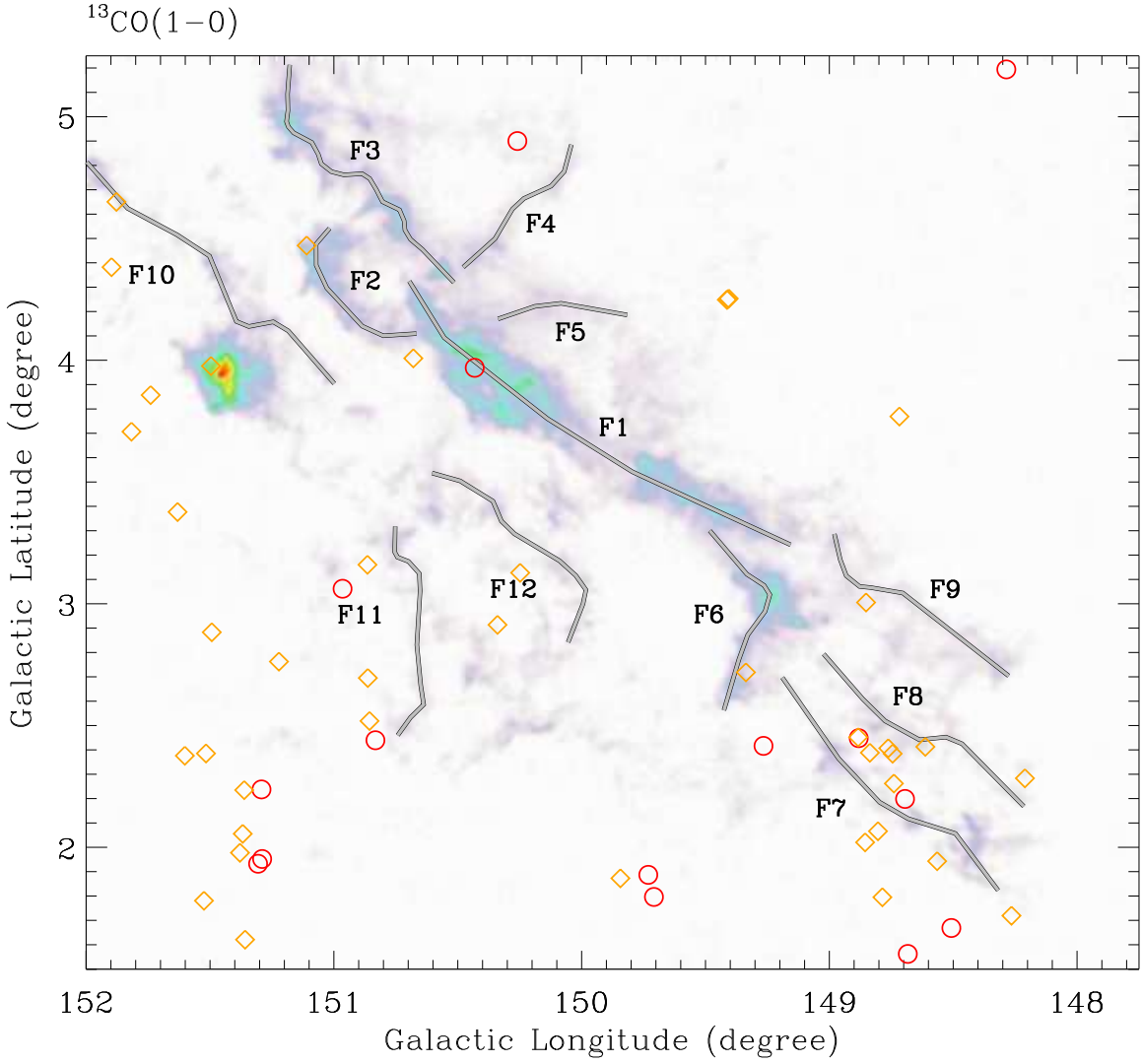}
\caption{Distribution of the identified YSOs with the background of $^{13}$CO integrated intensity map. The red circles are the Class~I objects and the orange diamonds are the Class~II objects. The solid white lines show the positions of identified filaments. The name of each filament is marked beside it.}
\label{fig:fig24}
\end{figure}

In Figure~\ref{fig:fig24}, we present the spatial distribution of the identified YSOs overlaid on the integrated intensity maps of $^{13}$CO. Most of the identified YSOs are not associated with the filaments. A group of YSOs is located in the southeast, far away from the filamentary structures. Another group of YSOs is located near the filaments F7 and F8 in the southwest. Two Class~II objects are found to be located in the northern area of F2 and the southern area of F6, respectively. One Class~I object is found in the central area of main filament F1.

In general, most virialized clumps and associated YSO candidates are found in the supercritical filaments, especially the virialized filaments (F1 and F2), indicating the existence of potential star-forming activities in these filaments. For the subcritical filaments F7 and F8, the small groups of virialized clumps and YSO candidates found in them may also indicate that this area contains star-forming activities, while for the rest of the subcritical filaments, almost no virialized clumps or associated YSO candidates are found in them, indicating that the star-forming activities may have not occurred in these filaments.

\section{SUMMARY}\label{sec:summary}
In this work, we present large-field mapping observations of the G150 region covering an area with $147.75\arcdeg \leq l \leq 152\arcdeg$ and $1.5\arcdeg \leq b \leq 5.25\arcdeg$ in the $J = 1-0$ emission line of CO isotopologues ($^{12}$CO, $^{13}$CO, and C$^{18}$O), using the PMO 13.7 m telescope. The CO gas spatial distribution and averaged spectra of the G150 region reveal three molecular gas layers with velocity ranges of $-$10.5 to $-$5~km~s$^{-1}$, $-$5 to $-$1.5~km~s$^{-1}$, and $-$0.5 to 6.5~km~s$^{-1}$, respectively. We focus on the third component ($-$0.5 -- 6.5~km~s$^{-1}$) in this work, which has significant $^{13}$CO and C$^{18}$O emission and shows conspicuous filamentary structures. The main results are listed below.

We identify 12 filaments (F1 to F12) in the $^{13}$CO image. The main filament F1 and the sub-filaments F2 to F9 in the surrounding area are associated with each other, while other sub-filaments F10 to F12 in the outer area are distributed approximately parallel to the main filament, forming the so-called ``ridge-nest'' structure together. The main filament in the ``ridge'' area is uni-directional with higher column density and the sub-filaments in the ``nest'' area are various directional with lower column density. We also identify four filaments (F1, F2, F3, and F6) in the C$^{18}$O image.

We extract the velocity structure along the length of each filament in both $^{13}$CO and C$^{18}$O data. The main filament and most sub-filaments have continuous velocity structures with slight velocity gradients in $^{13}$CO, and velocity structures in C$^{18}$O display discontinuous features, only concentrating on the densest parts. We derive the radial density profiles of the filaments and find that most profiles present Gaussian-like shapes in the inner parts. Based on the CO observations, the measured median excitation temperature, line width, length, width, and linear mass of the filaments are 9.28~K, 0.85~km~s$^{-1}$, 7.30~pc, 0.79~pc, and 17.92~$M_\sun$~pc$^{-1}$, respectively, assuming a distance of 400~pc.

After comparing LTE linear mass with critical linear mass ($M_{\rm crit}$ $\approx$ 18.56~$M_{\sun}$~pc$^{-1}$), only four filaments (F1, F2, F3, and F6), which are also detected in the C$^{18}$O image, are identified as thermally supercritical. We find that F1 and F2 have virial parameters smaller than two, suggesting that they are under virial equilibrium and could be under fragmentation, while F3 and F6 are close to virialized. The virial parameter of thermally subcritical filaments are much higher, indicating that they are not virialized and that they tend to be gravitationally unbound. We also find that all of the identified filaments have large turbulent motions with velocity dispersions much greater than the isothermal sound speed ($c_{\rm s}$ $\approx$ 0.2~km~s$^{-1}$).

We find in total 146 clumps in the $^{13}$CO data. Approximately 77$\%$ of the clumps are associated with the filaments, and 56$\%$ of the virialized clumps are associated with the thermally supercritical filaments. Based on the complementary IR data, one Class~I YSO and two Class~II YSOs are found to be located in the supercritical filaments F1, F2, and F6, respectively. The existence of virialized clumps and associated YSO candidates suggests that the thermally supercritical filaments, especially the virialized filaments (F1 and F2), may already have star-forming activities.

\acknowledgments
We are grateful to all the members of the Milky Way Imaging Scroll Painting CO line survey group, especially the staff of Qinghai Radio Station of PMO at Delingha for the support during the observations. We appreciate the anonymous referee for valuable comments and suggestions that helped to improve this paper. This work was supported by the National Natural Science Foundation of China (grants Nos. 11473069, 11233007, 11503086, and U1431231) and the Strategic Priority Research Program of the Chinese Academy of Sciences (grant No. XDB09000000). X.C. acknowledges the support of the Thousand Young Talents Program of China. This research has made use of the NASA/IPAC Infrared Science Archive, which is operated by the Jet Propulsion Laboratory, California Institute of Technology, under contract with the National Aeronautics and Space Administration.

\clearpage

\clearpage

\clearpage

\begin{deluxetable}{ccccccccc}
\tabletypesize{\scriptsize}
\tablewidth{0pt}
\tablecaption{Properties of Filaments\label{tab:tab1}}
\tablehead{
\colhead{Filament} & & \colhead{$T_{\rm ex}$} &  \colhead{$\Delta v$ ($^{13}$CO)} & \colhead{Mass ($^{13}$CO)} & \colhead{Length ($^{13}$CO)} & \colhead{Width ($^{13}$CO)} & \colhead{$M/l$ ($^{13}$CO)}    &  \colhead{Mass (C$^{18}$O)} \\
                   & & \colhead{(K)}          &  \colhead{(km s$^{-1}$)}          & \colhead{($M_\sun$)}       & \colhead{(pc)}               & \colhead{(pc)}              & \colhead{($M_\sun$ pc$^{-1}$)} &  \colhead{($M_\sun$)}         
}
\startdata
 F1 & &  10.14 &    1.00 & 849.82 &  13.41 &  1.80 &  63.37 &  200.28 \\
 F2 & &  10.13 &    0.73 & 201.52 &   5.03 &  1.03 &  40.05 &   50.62 \\
 F3 & &   9.92 &    0.98 & 360.63 &   8.76 &  0.94 &  41.19 &   16.14 \\
 F4 & &   7.90 &    0.80 &  74.10 &   4.86 &  0.72 &  15.25 &     $-$ \\
 F5 & &   8.74 &    0.86 &  67.26 &   3.75 &  1.17 &  17.92 &     $-$ \\
 F6 & &   9.32 &    0.91 & 230.60 &   6.17 &  0.84 &  37.39 &   21.56 \\
 F7 & &   8.39 &    0.70 & 136.96 &   8.93 &  0.78 &  15.34 &     $-$ \\
 F8 & &   9.29 &    0.64 &  77.25 &   7.51 &  0.75 &  10.29 &     $-$ \\
 F9 & &   9.28 &    0.73 & 128.48 &   6.84 &  0.81 &  18.80 &     $-$ \\
F10 & &   8.65 &    1.05 & 103.59 &  10.09 &  0.54 &  10.26 &     $-$ \\
F11 & &   8.06 &    0.79 &  81.03 &   6.69 &  0.49 &  12.11 &     $-$ \\
F12 & &   9.00 &    0.99 &  79.53 &   7.30 &  0.50 &  10.89 &     $-$ \\
\enddata
\tablecomments{Properties of the identified filaments, including excitation temperature (Column 2), line width in the $^{13}$CO emission (Column 3), LTE mass traced by the $^{13}$CO emission (Column 4), length in the $^{13}$CO emission (Column 5), linear mass traced by the $^{13}$CO emission (Column 6), and LTE mass traced by the C$^{18}$O emission (Column 7).}
\end{deluxetable}

\begin{deluxetable}{ccccccccc}
\tabletypesize{\scriptsize}
\tablewidth{0pt}
\tablecaption{Properties of Clumps in the G150 Region\label{tab:tab2}}
\tablehead{
\colhead{Clump} & & \colhead{Velocity}      & \colhead{Radius}    & \colhead{$T_{\rm ex}$} & \colhead{$\Delta v$}    & \colhead{$M_{\rm LTE}$} & \colhead{$M_{\rm Vir}$} & \colhead{$\alpha_{\rm Vir}$} \\
                & & \colhead{(km s$^{-1}$)} & \colhead{(pc)}      & \colhead{(K)}          & \colhead{(km s$^{-1}$)} & \colhead{($M_\sun$)}    & \colhead{($M_\sun$)}
}
\startdata 
MWISP G151.484$+$03.805  & &  2.25  &  0.076  &  10.31  &  1.04  &  1.67  &  17.25  &  10.35  \\
MWISP G148.434$+$02.117  & &  2.58  &  0.090  &  11.64  &  0.42  &  1.01  &   3.30  &   3.27  \\
MWISP G149.361$+$03.090  & &  3.07  &  0.089  &  10.56  &  0.37  &  0.91  &   2.56  &   2.83  \\
MWISP G151.109$+$04.446  & &  2.95  &  0.121  &  11.30  &  0.36  &  1.56  &   3.35  &   2.14  \\
MWISP G151.644$+$04.065  & &  1.60  &  0.074  &  10.58  &  0.32  &  0.66  &   1.58  &   2.39  \\
MWISP G148.502$+$02.453  & &  2.70  &  0.084  &  10.93  &  0.31  &  0.73  &   1.70  &   2.34  \\
MWISP G148.445$+$01.885  & &  1.91  &  0.067  &  12.71  &  0.39  &  0.60  &   2.11  &   3.50  \\
MWISP G150.454$+$02.688  & &  2.60  &  0.065  &   8.00  &  0.63  &  0.73  &   5.30  &   7.30  \\
MWISP G151.093$+$04.336  & &  3.85  &  0.080  &  10.94  &  0.40  &  0.66  &   2.69  &   4.06  \\
MWISP G151.018$+$03.931  & &  2.73  &  0.080  &  10.11  &  0.60  &  0.99  &   6.09  &   6.16  \\
MWISP G151.092$+$04.370  & &  3.06  &  0.094  &  11.21  &  0.19  &  0.73  &   0.75  &   1.02  \\
MWISP G148.352$+$01.851  & &  1.78  &  0.079  &   9.72  &  0.27  &  0.57  &   1.16  &   2.04  \\
MWISP G148.661$+$02.093  & &  1.59  &  0.106  &   5.63  &  0.58  &  1.48  &   7.37  &   4.98  \\
MWISP G151.859$+$04.639  & &  3.66  &  0.060  &  10.14  &  0.57  &  0.53  &   4.14  &   7.75  \\
MWISP G151.219$+$04.861  & &  3.44  &  0.118  &  10.59  &  0.59  &  1.79  &   8.50  &   4.76  \\
MWISP G151.244$+$04.169  & &  4.04  &  0.079  &   9.28  &  0.53  &  0.79  &   4.68  &   5.89  \\
MWISP G149.397$+$02.692  & &  2.31  &  0.087  &   9.55  &  0.36  &  0.72  &   2.42  &   3.33  \\
MWISP G149.010$+$02.937  & &  3.49  &  0.069  &   8.11  &  0.52  &  0.62  &   3.87  &   6.29  \\
MWISP G150.578$+$04.355  & &  3.43  &  0.096  &  12.59  &  0.57  &  1.22  &   6.45  &   5.31  \\
MWISP G148.492$+$01.996  & &  1.90  &  0.062  &   8.99  &  0.34  &  0.37  &   1.51  &   4.12  \\
MWISP G148.449$+$01.950  & &  2.42  &  0.057  &  10.04  &  0.28  &  0.36  &   0.90  &   2.52  \\
MWISP G151.484$+$04.770  & &  1.61  &  0.078  &  12.07  &  0.28  &  0.53  &   1.25  &   2.37  \\
MWISP G149.050$+$03.045  & &  3.53  &  0.072  &   7.36  &  0.39  &  0.53  &   2.25  &   4.22  \\
MWISP G151.511$+$04.405  & &  3.87  &  0.059  &   9.53  &  0.71  &  0.69  &   6.21  &   9.05  \\
MWISP G150.117$+$03.361  & &  1.56  &  0.045  &  10.28  &  0.82  &  0.53  &   6.40  &  12.14  \\
MWISP G150.975$+$04.188  & &  3.69  &  0.057  &  11.06  &  0.26  &  0.34  &   0.80  &   2.33  \\
MWISP G149.420$+$02.589  & &  0.88  &  0.071  &   8.98  &  0.42  &  0.52  &   2.65  &   5.08  \\
MWISP G149.435$+$02.645  & &  1.76  &  0.057  &   9.12  &  0.25  &  0.32  &   0.76  &   2.41  \\
MWISP G151.045$+$03.963  & &  2.42  &  0.092  &  10.32  &  0.37  &  0.75  &   2.61  &   3.46  \\
MWISP G148.486$+$02.076  & &  2.40  &  0.054  &   9.43  &  0.50  &  0.43  &   2.81  &   6.48  \\
MWISP G150.312$+$02.423  & &  2.90  &  0.070  &  10.85  &  0.31  &  0.43  &   1.41  &   3.27  \\
MWISP G148.302$+$02.234  & &  2.73  &  0.093  &  10.54  &  0.24  &  0.63  &   1.13  &   1.80  \\
MWISP G148.818$+$02.485  & &  2.42  &  0.078  &  10.87  &  0.52  &  0.76  &   4.41  &   5.82  \\
MWISP G149.376$+$02.737  & &  2.09  &  0.072  &  10.19  &  0.39  &  0.54  &   2.29  &   4.24  \\
MWISP G151.936$+$04.786  & &  3.87  &  0.060  &   9.40  &  0.23  &  0.32  &   0.64  &   1.98  \\
MWISP G151.929$+$04.730  & &  4.10  &  0.066  &   9.73  &  0.68  &  0.69  &   6.41  &   9.32  \\
MWISP G150.436$+$03.755  & &  3.42  &  0.076  &  11.16  &  0.33  &  0.53  &   1.72  &   3.25  \\
MWISP G148.743$+$02.392  & &  2.42  &  0.083  &  10.44  &  0.22  &  0.52  &   0.86  &   1.67  \\
MWISP G150.386$+$03.378  & &  1.58  &  0.110  &   8.66  &  0.17  &  0.71  &   0.69  &   0.97  \\
MWISP G150.837$+$03.254  & &  2.07  &  0.091  &   8.90  &  0.34  &  0.67  &   2.17  &   3.24  \\
MWISP G148.625$+$02.593  & &  3.09  &  0.068  &  11.12  &  0.26  &  0.42  &   0.98  &   2.32  \\
MWISP G150.888$+$04.504  & &  2.63  &  0.069  &   8.84  &  0.33  &  0.41  &   1.59  &   3.93  \\
MWISP G150.549$+$03.938  & &  3.42  &  0.047  &  11.05  &  0.39  &  0.27  &   1.46  &   5.42  \\
MWISP G149.027$+$02.328  & &  1.58  &  0.056  &   9.63  &  1.10  &  0.77  &  14.37  &  18.63  \\
MWISP G149.215$+$03.095  & &  3.73  &  0.078  &   9.05  &  0.41  &  0.51  &   2.69  &   5.32  \\
MWISP G150.455$+$02.735  & &  3.09  &  0.075  &   7.76  &  0.31  &  0.46  &   1.48  &   3.24  \\
MWISP G149.351$+$02.851  & &  2.13  &  0.085  &   9.17  &  0.41  &  0.62  &   3.01  &   4.84  \\
MWISP G150.738$+$03.213  & &  2.90  &  0.091  &   9.67  &  0.40  &  0.75  &   3.08  &   4.10  \\
MWISP G150.569$+$03.354  & &  2.25  &  0.070  &   9.42  &  0.29  &  0.38  &   1.22  &   3.21  \\
MWISP G149.276$+$03.042  & &  3.59  &  0.058  &  10.07  &  0.46  &  0.42  &   2.59  &   6.22  \\
MWISP G151.902$+$04.689  & &  4.04  &  0.055  &  12.30  &  0.39  &  0.38  &   1.73  &   4.57  \\
MWISP G150.068$+$03.414  & &  2.09  &  0.151  &  10.19  &  0.29  &  1.44  &   2.69  &   1.87  \\
MWISP G150.089$+$03.479  & &  2.40  &  0.073  &  10.75  &  0.32  &  0.48  &   1.56  &   3.28  \\
MWISP G151.603$+$04.022  & &  0.78  &  0.069  &  12.54  &  0.38  &  0.50  &   2.12  &   4.25  \\
MWISP G151.469$+$04.370  & &  2.08  &  0.071  &  11.00  &  0.24  &  0.42  &   0.83  &   2.00  \\
MWISP G149.332$+$02.995  & &  2.56  &  0.058  &  10.84  &  0.30  &  0.34  &   1.09  &   3.23  \\
MWISP G150.852$+$04.794  & &  3.12  &  0.068  &  12.18  &  0.35  &  0.45  &   1.77  &   3.96  \\
MWISP G150.696$+$03.297  & &  3.90  &  0.091  &   7.91  &  0.50  &  0.72  &   4.70  &   6.51  \\
MWISP G150.067$+$03.090  & &  2.24  &  0.056  &   8.26  &  0.56  &  0.42  &   3.69  &   8.81  \\
MWISP G150.610$+$04.446  & &  3.90  &  0.073  &  10.96  &  0.23  &  0.38  &   0.82  &   2.14  \\
MWISP G150.087$+$03.328  & &  1.60  &  0.059  &   9.71  &  0.60  &  0.49  &   4.48  &   9.19  \\
MWISP G148.994$+$02.402  & &  2.25  &  0.133  &   9.08  &  0.34  &  1.16  &   3.28  &   2.82  \\
MWISP G151.011$+$03.137  & &  0.89  &  0.056  &  10.52  &  0.26  &  0.27  &   0.78  &   2.87  \\
MWISP G150.692$+$04.252  & &  3.76  &  0.097  &  11.59  &  0.25  &  0.63  &   1.31  &   2.07  \\
MWISP G149.260$+$02.928  & &  3.22  &  0.089  &   7.30  &  0.22  &  0.53  &   0.93  &   1.77  \\
MWISP G151.154$+$03.571  & &  1.59  &  0.041  &   9.78  &  0.39  &  0.20  &   1.34  &   6.74  \\
MWISP G149.292$+$03.293  & &  3.99  &  0.122  &   9.67  &  0.25  &  0.83  &   1.60  &   1.91  \\
MWISP G150.736$+$03.513  & &  0.75  &  0.070  &   7.98  &  0.35  &  0.41  &   1.79  &   4.37  \\
MWISP G148.401$+$02.062  & &  2.33  &  0.074  &   8.50  &  0.58  &  0.54  &   5.13  &   9.49  \\
MWISP G148.410$+$02.009  & &  2.25  &  0.135  &  10.73  &  0.49  &  1.50  &   6.79  &   4.52  \\
MWISP G150.444$+$03.812  & &  3.44  &  0.100  &  10.52  &  0.38  &  0.77  &   2.97  &   3.84  \\
MWISP G149.444$+$02.611  & &  0.90  &  0.040  &   8.99  &  0.54  &  0.30  &   2.46  &   8.22  \\
MWISP G150.509$+$04.038  & &  3.43  &  0.016  &  11.12  &  0.31  &  0.03  &   0.32  &  10.05  \\
MWISP G149.340$+$02.797  & &  2.51  &  0.072  &  10.11  &  0.56  &  0.52  &   4.74  &   9.20  \\
MWISP G150.543$+$04.370  & &  5.38  &  0.093  &   7.47  &  0.49  &  0.75  &   4.76  &   6.32  \\
MWISP G149.291$+$02.969  & &  2.84  &  0.080  &  10.93  &  0.33  &  0.50  &   1.83  &   3.69  \\
MWISP G150.276$+$03.295  & &  1.66  &  0.052  &   9.28  &  0.32  &  0.26  &   1.12  &   4.37  \\
MWISP G151.036$+$04.721  & &  1.92  &  0.121  &   9.61  &  0.51  &  1.26  &   6.45  &   5.13  \\
MWISP G148.619$+$02.419  & &  3.07  &  0.087  &   8.29  &  0.44  &  0.63  &   3.44  &   5.48  \\
MWISP G150.552$+$03.979  & &  3.41  &  0.079  &  10.02  &  0.32  &  0.49  &   1.66  &   3.36  \\
MWISP G149.751$+$03.586  & &  3.75  &  0.114  &   6.26  &  0.34  &  0.83  &   2.71  &   3.27  \\
MWISP G149.327$+$03.077  & &  3.60  &  0.067  &  10.67  &  0.18  &  0.32  &   0.44  &   1.36  \\
MWISP G150.828$+$04.079  & &  3.54  &  0.085  &  11.17  &  0.24  &  0.49  &   1.01  &   2.06  \\
MWISP G150.046$+$03.180  & &  1.76  &  0.052  &   8.87  &  0.49  &  0.41  &   2.64  &   6.38  \\
MWISP G150.117$+$03.303  & &  2.17  &  0.043  &   9.77  &  0.25  &  0.16  &   0.54  &   3.31  \\
MWISP G150.092$+$03.009  & &  2.04  &  0.048  &   8.19  &  0.40  &  0.26  &   1.60  &   6.18  \\
MWISP G150.278$+$02.512  & &  3.39  &  0.054  &   7.00  &  0.64  &  0.43  &   4.57  &  10.73  \\
MWISP G149.117$+$02.932  & &  3.33  &  0.070  &   8.59  &  0.48  &  0.45  &   3.33  &   7.48  \\
MWISP G148.852$+$02.502  & &  2.92  &  0.075  &  10.73  &  0.21  &  0.40  &   0.67  &   1.68  \\
MWISP G150.070$+$02.437  & &  3.75  &  0.047  &   9.72  &  0.29  &  0.25  &   0.82  &   3.30  \\
MWISP G151.445$+$04.270  & &  3.73  &  0.068  &   6.65  &  0.62  &  0.52  &   5.43  &  10.47  \\
MWISP G150.986$+$04.521  & &  2.59  &  0.096  &   9.88  &  0.18  &  0.52  &   0.68  &   1.31  \\
MWISP G150.427$+$03.504  & &  1.25  &  0.066  &  10.31  &  0.58  &  0.54  &   4.74  &   8.74  \\
MWISP G150.719$+$04.604  & &  3.25  &  0.086  &  11.59  &  0.25  &  0.47  &   1.08  &   2.31  \\
MWISP G149.426$+$03.170  & &  2.91  &  0.098  &   9.39  &  0.25  &  0.56  &   1.33  &   2.37  \\
MWISP G149.126$+$03.327  & &  4.22  &  0.074  &   8.36  &  0.33  &  0.39  &   1.66  &   4.25  \\
MWISP G148.827$+$02.237  & &  2.06  &  0.048  &   8.84  &  0.27  &  0.18  &   0.73  &   4.05  \\
MWISP G148.310$+$02.001  & &  2.40  &  0.115  &  11.24  &  0.23  &  0.75  &   1.28  &   1.71  \\
MWISP G149.995$+$03.011  & &  2.22  &  0.051  &   7.57  &  0.29  &  0.20  &   0.88  &   4.34  \\
MWISP G150.870$+$04.746  & &  2.09  &  0.081  &  12.73  &  0.38  &  0.55  &   2.39  &   4.36  \\
MWISP G148.235$+$02.026  & &  2.42  &  0.071  &  10.25  &  0.33  &  0.42  &   1.61  &   3.86  \\
MWISP G150.076$+$03.153  & &  2.55  &  0.068  &   8.74  &  0.58  &  0.50  &   4.74  &   9.51  \\
MWISP G150.327$+$03.154  & &  1.89  &  0.080  &   8.73  &  0.28  &  0.40  &   1.35  &   3.42  \\
MWISP G149.019$+$02.759  & &  2.55  &  0.056  &  10.70  &  0.43  &  0.31  &   2.16  &   6.91  \\
MWISP G151.387$+$04.195  & &  3.58  &  0.072  &   9.48  &  0.76  &  0.72  &   8.66  &  12.09  \\
MWISP G149.168$+$02.936  & &  2.25  &  0.083  &  11.17  &  0.70  &  0.82  &   8.38  &  10.26  \\
MWISP G150.677$+$04.280  & &  4.40  &  0.088  &   9.95  &  0.26  &  0.46  &   1.21  &   2.62  \\
MWISP G150.137$+$02.911  & &  1.56  &  0.035  &   9.33  &  0.29  &  0.13  &   0.60  &   4.47  \\
MWISP G150.710$+$02.504  & &  1.08  &  0.068  &  10.87  &  0.30  &  0.36  &   1.33  &   3.64  \\
MWISP G148.543$+$02.552  & &  3.06  &  0.066  &  11.12  &  0.27  &  0.33  &   0.97  &   2.94  \\
MWISP G149.961$+$03.553  & &  3.22  &  0.078  &  10.60  &  0.18  &  0.33  &   0.50  &   1.51  \\
MWISP G148.401$+$01.868  & &  2.20  &  0.026  &  10.45  &  0.56  &  0.12  &   1.67  &  14.24  \\
MWISP G151.237$+$04.964  & &  3.43  &  0.081  &  10.37  &  0.36  &  0.46  &   2.20  &   4.83  \\
MWISP G151.419$+$03.838  & &  1.40  &  0.042  &   8.90  &  0.35  &  0.22  &   1.10  &   5.09  \\
MWISP G148.793$+$02.494  & &  2.92  &  0.024  &   9.22  &  0.29  &  0.06  &   0.40  &   6.58  \\
MWISP G150.484$+$03.953  & &  3.92  &  0.073  &   9.43  &  0.37  &  0.39  &   2.04  &   5.21  \\
MWISP G150.862$+$04.104  & &  3.53  &  0.054  &  10.31  &  0.17  &  0.22  &   0.32  &   1.46  \\
MWISP G150.661$+$04.204  & &  3.73  &  0.082  &  13.48  &  0.43  &  0.60  &   3.24  &   5.44  \\
MWISP G151.127$+$03.545  & &  1.92  &  0.050  &  10.73  &  0.17  &  0.19  &   0.31  &   1.62  \\
MWISP G150.561$+$03.662  & &  1.25  &  0.094  &  11.46  &  0.46  &  0.68  &   4.09  &   6.02  \\
MWISP G149.334$+$03.127  & &  3.72  &  0.107  &   8.47  &  0.24  &  0.58  &   1.34  &   2.31  \\
MWISP G150.552$+$04.070  & &  3.40  &  0.097  &  11.93  &  0.75  &  1.11  &  11.55  &  10.38  \\
MWISP G150.285$+$03.904  & &  2.58  &  0.140  &  12.53  &  0.16  &  0.91  &   0.72  &   0.79  \\
MWISP G150.319$+$03.337  & &  1.58  &  0.083  &   9.81  &  0.47  &  0.54  &   3.77  &   7.03  \\
MWISP G150.785$+$04.670  & &  3.40  &  0.085  &  10.49  &  0.67  &  0.73  &   8.06  &  11.11  \\
MWISP G149.402$+$02.661  & &  1.76  &  0.057  &   9.51  &  0.17  &  0.22  &   0.33  &   1.51  \\
MWISP G151.002$+$04.488  & &  3.23  &  0.073  &  11.64  &  0.29  &  0.37  &   1.27  &   3.47  \\
MWISP G149.102$+$02.919  & &  2.58  &  0.086  &  10.82  &  0.87  &  0.97  &  13.74  &  14.19  \\
MWISP G151.220$+$04.838  & &  3.90  &  0.066  &  10.29  &  0.44  &  0.34  &   2.60  &   7.54  \\
MWISP G148.717$+$02.362  & &  2.41  &  0.058  &   9.34  &  0.20  &  0.23  &   0.51  &   2.25  \\
MWISP G150.218$+$03.720  & &  3.57  &  0.079  &   9.56  &  0.34  &  0.41  &   1.96  &   4.84  \\
MWISP G149.326$+$03.369  & &  3.88  &  0.080  &   8.41  &  0.24  &  0.34  &   0.93  &   2.70  \\
MWISP G150.160$+$03.887  & &  3.89  &  0.140  &   8.75  &  0.16  &  0.78  &   0.72  &   0.93  \\
MWISP G149.167$+$03.303  & &  3.90  &  0.101  &  11.08  &  0.30  &  0.61  &   1.89  &   3.10  \\
MWISP G150.744$+$04.646  & &  3.41  &  0.071  &  10.27  &  0.32  &  0.36  &   1.50  &   4.15  \\
MWISP G149.276$+$03.120  & &  3.56  &  0.061  &   7.70  &  0.36  &  0.28  &   1.67  &   6.06  \\
MWISP G149.436$+$03.244  & &  3.58  &  0.078  &   9.80  &  0.26  &  0.35  &   1.08  &   3.12  \\
MWISP G148.377$+$02.076  & &  2.74  &  0.070  &   8.13  &  0.38  &  0.39  &   2.11  &   5.37  \\
MWISP G148.584$+$02.952  & &  2.39  &  0.064  &   9.00  &  0.32  &  0.28  &   1.41  &   5.07  \\
MWISP G151.961$+$04.747  & &  4.58  &  0.061  &   8.10  &  0.34  &  0.28  &   1.49  &   5.32  \\
MWISP G150.861$+$03.779  & &  3.07  &  0.068  &  10.15  &  0.53  &  0.44  &   3.99  &   9.14  \\
MWISP G150.727$+$04.503  & &  3.74  &  0.055  &  10.49  &  0.40  &  0.28  &   1.85  &   6.52  \\
MWISP G150.711$+$02.370  & &  0.91  &  0.056  &  12.06  &  0.39  &  0.29  &   1.80  &   6.16  \\
MWISP G148.644$+$02.541  & &  2.29  &  0.045  &   9.25  &  0.22  &  0.13  &   0.46  &   3.50  \\
MWISP G150.027$+$02.904  & &  1.57  &  0.054  &   7.26  &  0.56  &  0.32  &   3.57  &  11.30  \\
MWISP G151.052$+$04.937  & &  3.23  &  0.086  &  10.21  &  0.43  &  0.53  &   3.37  &   6.40  \\
\enddata
\tablecomments{Properties of the identified clumps. Column 1 is clump name, which contains Galactic longitude and latitude. Column 2 to 5 are LSR velocity, clump radius, excitation temperature, and line width. Column 6 to 8 are LTE mass, virial mass, and virial parameter.}
\end{deluxetable}

\begin{deluxetable}{ccccccccccc}
\centering
\tabletypesize{\scriptsize}
\tablewidth{0pt}
\setlength{\tabcolsep}{2.0pt}
\tablecaption{Infrared Photometric Magnitudes of YSO Candidates\label{tab:tab3}}
\tablehead{
\colhead{YSO} & \colhead{$l$}      & \colhead{$b$}      & \colhead{$J$ (1.25$\mu$m)} & \colhead{$H$ (1.65$\mu$m)} & \colhead{$K_{\rm s}$ (2.16$\mu$m)} & \colhead{$W1$ (3.4$\mu$m)} & \colhead{$W2$ (4.6$\mu$m)} & \colhead{$W3$ (12$\mu$m)} & \colhead{$W4$ (22$\mu$m)} & \colhead{Class} \\
              & \colhead{(arcdeg)} & \colhead{(arcdeg)} & \colhead{(mag)}            & \colhead{(mag)}            & \colhead{(mag)}                    & \colhead{(mag)}            & \colhead{(mag)}            & \colhead{(mag)}           & \colhead{(mag)}
}
\startdata
  1 & 148.68 & 1.56 & 14.95$\pm$0.04 & 13.82$\pm$0.04 & 12.79$\pm$0.03 & 11.43$\pm$0.02 & 10.24$\pm$0.02 &  7.57$\pm$0.02 &  5.33$\pm$0.04 & I  \\
  2 & 148.51 & 1.67 &      $-$       &      $-$       &      $-$       & 14.25$\pm$0.03 & 12.28$\pm$0.02 &  8.52$\pm$0.02 &  6.54$\pm$0.06 & I  \\
  3 & 148.69 & 2.20 & 17.21$\pm$0.20 & 15.24$\pm$0.10 & 13.41$\pm$0.04 & 11.63$\pm$0.02 &  9.91$\pm$0.02 &  6.82$\pm$0.02 &  4.21$\pm$0.02 & I  \\
  4 & 148.88 & 2.45 & 17.11$\pm$0.17 & 16.02$\pm$0.15 & 15.05$\pm$0.10 & 13.26$\pm$0.04 & 11.76$\pm$0.03 &  7.67$\pm$0.02 &  5.49$\pm$0.04 & I  \\
  5 & 148.29 & 5.19 & 16.69$\pm$0.16 & 16.20$\pm$0.23 & 14.74$\pm$0.10 & 12.83$\pm$0.02 & 11.36$\pm$0.02 &  8.28$\pm$0.02 &  6.00$\pm$0.05 & I  \\
  6 & 149.71 & 1.80 & 15.93$\pm$0.08 & 14.71$\pm$0.07 & 13.95$\pm$0.06 & 12.47$\pm$0.02 & 11.29$\pm$0.02 &  8.30$\pm$0.02 &  5.86$\pm$0.04 & I  \\
  7 & 149.73 & 1.89 &      $-$       &      $-$       &      $-$       & 15.90$\pm$0.06 & 12.92$\pm$0.03 &  9.10$\pm$0.03 &  6.22$\pm$0.06 & I  \\
  8 & 149.27 & 2.42 & 14.94$\pm$0.07 & 13.76$\pm$0.07 & 13.00$\pm$0.04 & 11.88$\pm$0.03 & 10.64$\pm$0.02 &  6.14$\pm$0.01 &  3.13$\pm$0.02 & I  \\
  9 & 150.83 & 2.44 &      $-$       &      $-$       &      $-$       & 11.93$\pm$0.03 & 10.77$\pm$0.02 &  8.24$\pm$0.03 &  5.77$\pm$0.06 & I  \\
 10 & 150.97 & 3.06 & 16.27$\pm$0.09 & 14.78$\pm$0.07 & 13.35$\pm$0.04 & 11.80$\pm$0.02 & 10.76$\pm$0.02 &  7.94$\pm$0.02 &  6.11$\pm$0.06 & I  \\
 11 & 150.26 & 4.90 & 16.35$\pm$0.11 & 15.30$\pm$0.09 & 14.56$\pm$0.08 & 12.90$\pm$0.02 & 11.77$\pm$0.02 &  8.97$\pm$0.03 &  7.01$\pm$0.09 & I  \\
 12 & 151.31 & 1.93 &      17.77     & 15.44$\pm$0.12 & 13.64$\pm$0.05 & 10.38$\pm$0.02 &  9.04$\pm$0.02 &  6.65$\pm$0.03 &  4.64$\pm$0.04 & I  \\
 13 & 151.29 & 2.24 & 16.32$\pm$0.11 & 14.49$\pm$0.07 & 13.43$\pm$0.04 & 11.89$\pm$0.02 & 10.68$\pm$0.02 &  7.69$\pm$0.02 &  5.17$\pm$0.03 & I  \\
 14 & 150.43 & 3.97 &      18.83     & 14.68$\pm$0.05 & 12.58$\pm$0.02 & 12.06$\pm$0.02 & 10.93$\pm$0.02 & 11.51$\pm$0.20 &      9.22      & I  \\
 15 & 151.29 & 1.95 & 15.11$\pm$0.05 & 13.52$\pm$0.04 & 12.50$\pm$0.03 & 11.04$\pm$0.02 &  9.92$\pm$0.02 &  6.70$\pm$0.03 &  3.94$\pm$0.04 & I  \\
 16 & 148.79 & 1.79 & 13.46$\pm$0.03 & 13.05$\pm$0.03 & 12.76$\pm$0.03 & 12.33$\pm$0.03 & 12.02$\pm$0.02 & 10.71$\pm$0.10 &  8.77$\pm$0.46 & II \\
 17 & 148.56 & 1.94 & 14.94$\pm$0.04 & 14.21$\pm$0.04 & 13.85$\pm$0.04 & 12.89$\pm$0.03 & 12.15$\pm$0.02 &  9.28$\pm$0.06 &      8.13      & II \\
 18 & 148.26 & 1.72 & 14.72$\pm$0.04 & 14.11$\pm$0.05 & 13.63$\pm$0.04 & 12.52$\pm$0.02 & 12.15$\pm$0.02 & 10.75$\pm$0.11 &  8.59$\pm$0.35 & II \\
 19 & 148.86 & 2.02 & 15.23$\pm$0.05 & 13.35$\pm$0.04 & 11.90$\pm$0.02 &  8.65$\pm$0.02 &  7.93$\pm$0.02 &  5.24$\pm$0.01 &  2.49$\pm$0.02 & II \\
 20 & 148.80 & 2.07 & 15.90$\pm$0.09 & 14.63$\pm$0.09 & 13.87$\pm$0.08 & 12.38$\pm$0.02 & 11.46$\pm$0.02 &  9.18$\pm$0.06 &  6.69$\pm$0.08 & II \\
 21 & 148.74 & 2.26 & 16.39$\pm$0.11 & 14.68$\pm$0.07 & 13.87$\pm$0.05 & 12.81$\pm$0.03 & 12.17$\pm$0.02 & 10.46$\pm$0.11 &  8.24$\pm$0.34 & II \\
 22 & 148.74 & 2.38 & 15.84$\pm$0.07 & 14.56$\pm$0.07 & 13.79$\pm$0.04 & 12.83$\pm$0.03 & 11.86$\pm$0.02 &  9.05$\pm$0.05 &  6.61$\pm$0.08 & II \\
 23 & 148.84 & 2.39 & 14.03$\pm$0.04 & 13.29$\pm$0.05 & 12.70$\pm$0.03 & 11.59$\pm$0.02 & 11.09$\pm$0.02 &  9.47$\pm$0.04 &  7.92$\pm$0.20 & II \\
 24 & 148.88 & 2.45 & 15.45$\pm$0.07 & 14.48$\pm$0.09 & 13.55$\pm$0.05 & 12.10$\pm$0.02 & 11.38$\pm$0.02 &  8.83$\pm$0.03 &  6.17$\pm$0.05 & II \\
 25 & 148.76 & 2.41 & 16.09$\pm$0.07 & 14.76$\pm$0.06 & 13.73$\pm$0.04 & 12.66$\pm$0.02 & 11.95$\pm$0.02 &  9.33$\pm$0.04 &  6.95$\pm$0.09 & II \\
 26 & 148.61 & 2.41 & 14.75$\pm$0.04 & 12.46$\pm$0.04 & 10.80$\pm$0.02 &  8.91$\pm$0.02 &  7.87$\pm$0.02 &  5.66$\pm$0.01 &  3.55$\pm$0.02 & II \\
 27 & 148.21 & 2.28 & 16.70$\pm$0.13 & 14.80$\pm$0.07 & 13.31$\pm$0.04 &  9.13$\pm$0.02 &  8.17$\pm$0.02 &  5.54$\pm$0.01 &  3.56$\pm$0.02 & II \\
 28 & 148.85 & 3.01 & 14.10$\pm$0.03 & 13.08$\pm$0.03 & 12.20$\pm$0.02 & 11.01$\pm$0.02 & 10.35$\pm$0.02 &  8.46$\pm$0.02 &  6.64$\pm$0.07 & II \\
 29 & 148.72 & 3.77 & 15.39$\pm$0.17 & 14.56$\pm$0.10 & 13.63$\pm$0.07 & 12.18$\pm$0.03 & 11.60$\pm$0.02 &  9.05$\pm$0.03 &  6.93$\pm$0.10 & II \\
 30 & 149.84 & 1.87 & 14.87$\pm$0.05 & 14.05$\pm$0.05 & 13.07$\pm$0.04 & 11.71$\pm$0.02 & 10.71$\pm$0.02 &  7.99$\pm$0.02 &  5.86$\pm$0.06 & II \\
 31 & 149.34 & 2.72 & 13.36$\pm$0.02 & 12.66$\pm$0.02 & 12.46$\pm$0.03 & 12.08$\pm$0.02 & 11.76$\pm$0.02 & 10.46$\pm$0.09 &      9.17      & II \\
 32 & 149.41 & 4.25 & 14.34$\pm$0.04 & 13.27$\pm$0.03 & 12.27$\pm$0.02 & 10.96$\pm$0.02 & 10.24$\pm$0.02 &  7.99$\pm$0.02 &  6.26$\pm$0.06 & II \\
 33 & 149.41 & 4.25 & 16.31$\pm$0.10 & 14.23$\pm$0.05 & 12.95$\pm$0.04 & 11.30$\pm$0.02 & 10.34$\pm$0.02 &  7.96$\pm$0.02 &  5.59$\pm$0.04 & II \\
 34 & 149.41 & 4.25 &      14.85     &      13.92     & 14.07$\pm$0.10 & 12.42$\pm$0.03 & 11.74$\pm$0.03 &  9.57$\pm$0.05 &  6.67$\pm$0.09 & II \\
 35 & 150.86 & 2.52 & 13.54$\pm$0.04 & 12.44$\pm$0.04 & 11.81$\pm$0.03 & 10.92$\pm$0.02 & 10.32$\pm$0.02 &  8.01$\pm$0.03 &  5.68$\pm$0.06 & II \\
 36 & 150.86 & 2.70 & 16.76$\pm$0.16 & 15.66$\pm$0.15 & 14.53$\pm$0.09 & 12.35$\pm$0.02 & 11.21$\pm$0.02 &  9.08$\pm$0.04 &  7.00$\pm$0.09 & II \\
 37 & 150.34 & 2.91 & 13.05$\pm$0.03 & 11.63$\pm$0.03 & 10.53$\pm$0.02 &  9.31$\pm$0.02 &  8.27$\pm$0.02 &  5.83$\pm$0.02 &  4.20$\pm$0.03 & II \\
 38 & 150.86 & 3.16 & 12.74$\pm$0.03 & 12.44$\pm$0.02 & 12.12$\pm$0.02 & 11.88$\pm$0.03 & 11.58$\pm$0.02 & 10.20$\pm$0.07 &  8.37$\pm$0.32 & II \\
 39 & 150.25 & 3.13 & 13.37$\pm$0.03 & 12.48$\pm$0.03 & 12.10$\pm$0.02 & 11.61$\pm$0.02 & 11.33$\pm$0.02 &  8.78$\pm$0.03 &  6.43$\pm$0.06 & II \\
 40 & 150.68 & 4.01 & 13.74$\pm$0.03 & 12.79$\pm$0.03 & 12.40$\pm$0.02 & 11.98$\pm$0.02 & 11.42$\pm$0.02 &  9.53$\pm$0.05 &  8.65$\pm$0.48 & II \\
 41 & 151.52 & 1.78 & 15.86$\pm$0.09 & 14.59$\pm$0.09 & 13.42$\pm$0.05 & 11.98$\pm$0.02 & 11.02$\pm$0.02 &  8.44$\pm$0.03 &  6.24$\pm$0.06 & II \\
 42 & 151.38 & 1.98 & 14.43$\pm$0.04 & 13.60$\pm$0.04 & 12.90$\pm$0.03 & 11.84$\pm$0.02 & 11.16$\pm$0.02 &  8.38$\pm$0.02 &  6.48$\pm$0.06 & II \\
 43 & 151.36 & 1.62 & 15.00$\pm$0.06 & 13.84$\pm$0.05 & 12.83$\pm$0.03 & 11.66$\pm$0.02 & 10.69$\pm$0.02 &  8.18$\pm$0.03 &  5.39$\pm$0.04 & II \\
 44 & 151.60 & 2.38 & 13.66$\pm$0.04 & 12.88$\pm$0.05 & 12.37$\pm$0.04 & 11.49$\pm$0.02 & 10.84$\pm$0.02 &  8.35$\pm$0.03 &  6.75$\pm$0.08 & II \\
 45 & 151.52 & 2.39 & 14.32$\pm$0.04 & 13.17$\pm$0.03 & 12.35$\pm$0.02 & 11.35$\pm$0.02 & 10.69$\pm$0.02 &  8.15$\pm$0.02 &  6.17$\pm$0.05 & II \\
 46 & 151.49 & 2.88 & 14.80$\pm$0.03 & 13.71$\pm$0.03 & 13.07$\pm$0.03 & 12.16$\pm$0.02 & 11.73$\pm$0.02 & 10.36$\pm$0.07 &  8.38$\pm$0.28 & II \\
 47 & 151.37 & 2.06 & 15.28$\pm$0.05 & 14.33$\pm$0.06 & 13.57$\pm$0.04 & 12.56$\pm$0.02 & 12.07$\pm$0.02 &  9.10$\pm$0.04 &  6.46$\pm$0.07 & II \\
 48 & 151.36 & 2.23 & 14.89$\pm$0.05 & 14.21$\pm$0.05 & 13.68$\pm$0.05 & 12.85$\pm$0.02 & 12.37$\pm$0.02 & 10.62$\pm$0.09 &  7.84$\pm$0.19 & II \\
 49 & 151.22 & 2.76 & 15.26$\pm$0.07 & 14.57$\pm$0.05 & 13.78$\pm$0.04 & 12.77$\pm$0.02 & 12.19$\pm$0.02 &  9.38$\pm$0.04 &  7.47$\pm$0.13 & II \\
 50 & 151.82 & 3.71 & 14.98$\pm$0.05 & 13.92$\pm$0.04 & 13.10$\pm$0.04 & 12.22$\pm$0.02 & 11.71$\pm$0.02 & 10.35$\pm$0.08 &  8.14$\pm$0.26 & II \\
 51 & 151.63 & 3.38 & 15.67$\pm$0.07 & 14.72$\pm$0.06 & 13.90$\pm$0.05 & 13.00$\pm$0.02 & 12.15$\pm$0.03 &  9.42$\pm$0.04 &  6.42$\pm$0.06 & II \\
 52 & 151.74 & 3.86 & 14.64$\pm$0.04 & 13.80$\pm$0.04 & 13.32$\pm$0.04 & 12.60$\pm$0.03 & 11.98$\pm$0.02 & 10.11$\pm$0.07 &  7.96$\pm$0.28 & II \\
 53 & 151.50 & 3.98 & 13.93$\pm$0.03 & 12.48$\pm$0.02 & 11.83$\pm$0.02 & 11.02$\pm$0.02 & 10.46$\pm$0.02 &  7.55$\pm$0.02 &  5.26$\pm$0.03 & II \\
 54 & 151.90 & 4.38 & 14.28$\pm$0.03 & 13.34$\pm$0.03 & 12.75$\pm$0.03 & 12.64$\pm$0.03 & 12.18$\pm$0.03 & 10.37$\pm$0.07 &  8.19$\pm$0.22 & II \\
 55 & 151.88 & 4.65 & 16.34$\pm$0.12 & 15.17$\pm$0.09 & 13.96$\pm$0.05 & 12.13$\pm$0.02 & 11.09$\pm$0.02 &  8.59$\pm$0.03 &  6.15$\pm$0.06 & II \\
 56 & 151.92 & 5.23 &      16.62     &      16.18     & 15.17$\pm$0.13 & 12.51$\pm$0.02 & 11.37$\pm$0.02 &  9.53$\pm$0.04 &  7.64$\pm$0.15 & II \\
 57 & 151.11 & 4.47 &      16.89     & 14.31$\pm$0.06 & 13.03$\pm$0.03 & 12.67$\pm$0.02 & 12.04$\pm$0.02 &      12.27     &      9.02      & II \\
\enddata
\tablecomments{Infrared photometric magnitudes of the identified YSO candidates. Columns 2 and 3 list the Galactic longitudes and latitudes of YSOs. Columns 4 to 6 list the photometric magnitudes of 2MASS $J$, $H$, and $K_{\rm s}$ bands. Columns 7 to 10 list the photometric magnitudes of WISE $W1$, $W2$, $W3$, and $W4$ bands. Column 11 lists the classification of YSOs. For the photometric magnitudes without uncertainties, the values presented here are the upper limits.}
\end{deluxetable}

\end{document}